\documentclass[authoryear,11pt]{elsarticle}

\usepackage{graphicx} %
\usepackage[dvipsnames]{xcolor}
\usepackage[left=1in,right=1in,top=1in,bottom=1in]{geometry}
\usepackage{amsmath}
\usepackage{amssymb}
\usepackage{amsfonts}
\usepackage{bm}
\usepackage{float}
\usepackage{tikz,tikz-cd}
\tikzcdset{arrows={line width=rule_thickness},arrow style=math font}

\usepackage{soul}
\usepackage{mathrsfs}
\usepackage{breqn}
\usepackage{hyperref}
\usepackage{subfig}
\usepackage{comment}
\usepackage{cancel}
\usepackage[section]{placeins}

\newcommand{\pos}{\ensuremath{\mathcal{I}}}
\newcommand{\posp}{\ensuremath{\mathcal{I}'}}
\newcommand{\vac}{\ensuremath{\mathcal{V}}}
\newcommand{\vacp}{\ensuremath{\mathcal{V}'}}

\usepackage{lineno}

\affiliation[1]{organization={Johns Hopkins University},addressline={ Department of Applied Mathematics and Statistics}, city={Baltimore}, state={MD}, postcode={21218}, country={USA}}
\affiliation[2]{organization={National Institute of Standards and Technology}, addressline={Information Technology Laboratory}, city={Gaithersburg}, state={MD}, postcode={20899}, country={USA}}
\affiliation[3]{organization={George Mason University}, addressline={Department of Mathematical Sciences}, city={Fairfax}, state={VA}, postcode={22030}, country={USA}}

\author[1,2]{Prajakta Bedekar\fnref{contrib}} 
\author[2,3]{Rayanne A. Luke\fnref{contrib}\corref{corauth}} \author[2]{Anthony J. Kearsley}
\fntext[contrib]{These authors contributed equally to this work.}
\cortext[corauth]{Corresponding author at: George Mason University, Department of Mathematical Sciences, Fairfax, VA, 22030, USA. E-mail address: rluke@gmu.edu (R.A. Luke).}

\hypersetup{
  pdftitle={Prevalence estimation methods for time-dependent antibody kinetics of infected and vaccinated individuals: a graph-theoretic approach},
  pdfauthor={Bedekar, Luke, and Kearsley}
}

\allowdisplaybreaks 

\begin{document}

\begin{frontmatter}

\title{Prevalence estimation methods for time-dependent antibody kinetics of infected and vaccinated individuals: a graph-theoretic approach}

\begin{abstract}
Immune events such as infection, vaccination, and a combination of the two result in distinct time-dependent antibody responses in affected individuals. These responses and event prevalences combine non-trivially to govern antibody levels sampled from a population. Time-dependence and disease prevalence pose considerable modeling challenges that need to be addressed to provide a rigorous mathematical underpinning of the underlying biology. We propose a time-inhomogeneous Markov chain model for event-to-event transitions coupled with a probabilistic framework for antibody kinetics and demonstrate its use in a setting in which individuals can be infected or vaccinated but not both. We prove the equivalency of this approach to the framework developed in our previous work. Synthetic data are used to demonstrate the modeling process and conduct prevalence estimation via transition probability matrices. This approach is ideal to model sequences of infections and vaccinations, or personal trajectories in a population, making it an important first step towards a mathematical characterization of reinfection, vaccination boosting, and cross-events of infection after vaccination or vice versa.

\end{abstract}

\begin{keyword}

Antibody testing \sep diagnostics \sep graph-theoretic models \sep prevalence estimation \sep time-dependence 

\end{keyword}

\end{frontmatter}

\section{Introduction}

As a disease becomes endemic and vaccination widespread, the ability to characterize the antibody response across time and events can become as important as the capacity to track epidemiological trends. Analysis of antibody testing can characterize the immune response to infection or vaccination and provides information with which to make population-level decisions \citep{caini2020meta,peeling2020serology}. 
Accurate interpretation of an antibody measurement randomly sampled from a population should depend on the prevalence at the time of measurement, whether the sample was obtained from a vaccinated or infected individual, and when the measurement was taken. These effects can be seen in biological literature, and have been separately modeled in cohesive probabilistic frameworks \citep{patrone2021classification,  bedekar2022prevalence,
bottcher2022statistical,patrone2022holdout,  luke2023optimal}. However, existing methods are not equipped to consider both transmission dynamics and time-dependent antibody response in a multiclass setting.

To overcome these real-world obstacles, we must consider mathematical models that bridge multiple scales by simultaneously considering the effects of (i) prevalence, (ii) more than two classes, and (iii) time-dependence. A shortcoming of canonical models for communicable diseases, such as susceptible-infected-recovered (SIR) models or statistical regression models, is their inability to track population-level antibody responses across time, even though many capture global transmission dynamics of viral infections by using a proxy for immune response  \citep[e.g.,][]{d2020assessment,mcmahon2020reinfection,quick2021regression,roberto2021sars}. In one example, \cite{dick2021covid} built an age-structured SIR-type model of boosting and waning immunity to severe acute respiratory syndrome coronavirus 2 (SARS-CoV-2) infection. Their model estimates prevalence as the total population with some ``immunity''  (not defined). To address viral evolution and antibody response, within-host differential equation models have been created, studying questions such as the protection time of antibodies \citep[e.g.,][]{wodarz2005mathematical,hernandez2020host,depillis2023mathematical,xu2023novel}; such models ignore population-level trends.  Finally, as an example of an approach that addresses time-dependence and antibody levels, \cite{hay2019characterising} applied Markov chain Monte Carlo methods to  influenza infections or vaccinations in ferrets.
None of these works addressed all three effects (i, ii, and iii) simultaneously, nor their complex multi-scale interactions.

There is a rich literature on longitudinal immune response dynamics \citep[e.g.,][]{diep2023successive,guo2023durability,liu2023persistence} and many  seroprevalence studies have been conducted \citep[e.g.,][]{osborne2000ten,pollan2020prevalence,bajema2021estimated}; a mathematical framework is needed to analyze measurements and make decisions. This is no easy feat; time-dependence, prevalence, and multiple classes combine non-trivially and  pose considerable modeling challenges. In prior work we have addressed questions of time-dependence and multiclass situations separately via probabilistic modeling \citep{bedekar2022prevalence, luke2023optimal}; both included prevalence estimation schemes.
 These works constructed mathematical machinery to calculate the probabilities of events of interest, such as single infections, but not a way to track transitions from one event to another or estimate an individual's antibody levels associated with a sequence of events.   To fully address the task of modeling real-life situations, a time-inhomogeneous Markov chain model for event-to-event transitions can be used to extend the probabilistic framework already in place. 

In this paper, we combine our prior work  \citep{bedekar2022prevalence, luke2023optimal} into a probabilistic model for the time-dependent antibody kinetics of the situation in  which an individual either gets infected or vaccinated and then stays in that class (Section \ref{sec:methods}).   
Using methods mirroring \cite{bedekar2022prevalence} and \cite{luke2023optimal}, we 
develop a prevalence estimation scheme for na{\"i}ve,  infected, or vaccinated samples. We then present the same problem through the lens of a time-inhomogeneous Markov chain model
(Section \ref{sec:network_approach}); such a formulation facilitates generalization. We demonstrate the equivalence of our extension of prior work and Markov chain frameworks by defining the transition probabilities through the 
class incidences and prevalences on a given time step. In addition, we develop a transition probability matrix estimation framework that allows for an equivalent method of prevalence estimation.  We validate our methods using synthetic data based on SARS-CoV-2 serological measurements \citep{abela2021multifactorial,congrave2022twelve}\footnote{Certain commercial equipment, instruments, software, or material are identified in this paper in order to specify the experimental procedure adequately. Such identification is not intended to imply recommendation or endorsement by the National Institute of Standards and Technology, nor is it intended to imply that the materials or equipment identified are necessarily the best available for the purpose.} in Section \ref{sec:example}. The discussion includes further analysis of prevalence estimation, comparisons to other approaches, limitations, and extensions (Section \ref{sec:disc}). We present this work 
to bridge the gap to the most general model in which reinfections, revaccinations, and cross events will be allowed. 

\section{Notation}

\noindent Below is a summary of baseline terminology and descriptions of terms as they pertain to our work. 

\subsection{Definitions from applied diagnostics}

\begin{itemize}

\item The na{\"i}ve class consists of individuals who have no history of infection or vaccination. In a binary classification setting, such individuals are often referred to as ‘negative’.

\item The infected class consists of individuals who have been acutely or previously infected but who are unvaccinated. In a binary classification setting, such individuals are often referred to as ‘positive’. 

\item The vaccinated class consists of individuals with a  new or prior inoculation against a disease without a prior infection.

\item Incidence refers to the fraction of new infections or vaccinations in the total population during a given time step \citep{bouter2023textbook}. We define an infection incidence and a vaccination incidence.

\item  A class prevalence during a given time step after the emergence of a disease is the fraction of individuals in the population in that class on that time step and is the sum of the incidences over all previous time steps.

\item Training data correspond to sample antibody measurements from individuals for whom the true classes are known. Typically, such data are used to construct conditional probability models.

\item Test data correspond to sample antibody measurements from individuals for whom the true classes are unknown or assumed to be unknown for validation purposes. Typically, a prevalence estimation procedure is applied to such data.

\item Personal timeline refers to the duration since infection or vaccination for an individual.

\item Absolute timeline denotes time relative to the emergence of the disease.

\end{itemize}

\subsection{Notation specific to this paper}
\begin{itemize}
\item Antibody measurement is denoted by vector $\bm{r}$. The set $\Omega$ denotes the entire measurement space.
\item The prevalence for each class is denoted by $q_J$ and the incidence by $f_J$, with $J = N, I, V$ denoting the class as na{\"i}ve, infected, or vaccinated. These are functions of time.
\item The use of the symbol $\ \widehat{}\ $ denotes an estimated quantity.

\end{itemize}

\section{Multiclass extension of existing time-dependent theory}

\label{sec:methods}
In this section, we combine our prior work \citep{bedekar2022prevalence, luke2023optimal} to arrive at a probabilistic model for the time-dependent antibody kinetics of the situation in  which an individual either gets infected or vaccinated, and then stays in that class. Here, the focus lies on explicitly enumerating all the ways in which, for example, an individual could have been infected by a certain time period. This will be contrasted with a graph theoretic approach in Section \ref{sec:network_approach} where the focus is on keeping track of transitions in the population every time period.

For both Sections \ref{sec:methods} and \ref{sec:network_approach}, the following holds true. A blood or saliva sample from an individual is measured to obtain an antibody measurement $\bm{r}$, a vector in some compact domain $\Omega \subset \mathbb{R}^n$. The boundaries of $\Omega$ are governed by the measurement range of the instrument used. We generally use $t$ to indicate time in the personal timeline, which is the duration since infection or vaccination for an individual. We generally use $T$ to denote time in the absolute timeline in the emergence of the disease. We consider time to be discrete in this manuscript, as antibody measurements are generally reported at regular time intervals.

\subsection{Probability models}
\label{sec:prob_models}

The antibody response to infection or vaccination is time dependent, but that of an immuno-na{\"i}ve individual is not. This is a basic assumption of our probability models.
Let $N(\bm{r}) = \text{Prob}(\bm{r} | \text{Na{\"i}ve})$ denote the probability density that a sample yields measurement $\bm{r}$ given that the true underlying class is na{\"i}ve. Here we use Prob to denote probability density.
Let $I(\bm{r}, T) = \text{Prob}(\bm{r}, T | \text{Infected})$ give the probability density that a sample yields measurement $\bm{r}$ on time step $T$ of the absolute timeline given that the true underlying class is infected. 
$V(\bm{r}, T)$ is similarly defined for the vaccinated class. Note that in contrast to an SIR framework, we have no recovered class;  the infected class consists of individuals whose symptoms may have subsided but their antibody response is still determined by the infection event. For our three classes, we assume

\begin{equation}
N:\ \Omega \rightarrow \mathbb{R}^+,\ \ \ I,V: \Omega \times \{0,1,2,\cdots \} \rightarrow \mathbb{R}^+.
\end{equation}
We consider time to be discretized such that one time step is of length $dt$ days, as  information is reported, e.g., once per day or as seven-day averages of new caseloads.

Each class has an associated prevalence: $q_N(T), q_I(T),$ and $ q_V(T)$, denoting the fraction of the population in each class at time step $T$. Prevalence quantifies the total fraction of the population incident into that class so far and thus takes values in the range $[0,1]$. Since we assume that reinfection, revaccination, and infection after vaccination or vice versa do not occur, 
 once someone is infected, they move into the infected class and stay there; similarly for vaccination.  As a result, $q_I$ and $q_V$ increase over time, and $q_N$ decreases over time. 

The probabilities above combine to form the measurement density $Q(\bm{r}, T)$ that a sample collected on time step $T$ has antibody level $\bm{r}$. The law of total probability gives

\begin{equation}
    Q(\bm{r},T) =  q_N(T) N(\bm{r}) + q_I(T) I (\bm{r}, T)  + q_V(T) V(\bm{r},T).
    \label{eq:Q}
\end{equation}
We want to construct $I(\bm{r}, T)$, which is naturally composed of the probabilities of being infected on different time steps before time step $T$. Via the law of total probability, following \cite{bedekar2022prevalence}, we find

\begin{equation}
I(\bm{r}, T) = \sum_{t = 0}^T \text{Prob}(\bm{r}, T, \text{infected on time step } t). 
\label{eq:I_prob}
\end{equation}
Note that this conditional probability density can be determined in this straightforward way because the set of collectively exhausted events are defined by the date of infection, since we assume this occurs once and only once.
Let $R$ denote the probability density of observing a measurement $\bm{r}$, $t$ time steps after infection. We then have

\begin{equation}
I(\bm{r},T) = \sum_{t = 0}^T R(\bm{r},t) \frac{f_I(T - t)}{q_I(T)} = \sum_{t = 0}^T R(\bm{r}, T-t) \frac{f_I(t)}{q_I(T)},
\label{eq:P}
\end{equation}
where $f_I(T)$ describes the  infection incidence, or fraction of the total population that is infected on time step $T$ of the absolute timeline. Define $V(\bm{r},t)$ and $f_V$ similarly, where $W$ is analogous to $R$:

\begin{equation}
V(\bm{r},T) = \sum_{t = 0}^T W(\bm{r},t) \frac{f_V(T - t)}{q_V(T)} = \sum_{t = 0}^T W(\bm{r}, T-t) \frac{f_V(t)}{q_V(T)}.
\label{eq:V}
\end{equation}
Due to the assumptions stated earlier, $f_I, f_V \geq 0$. The infection and vaccination incidences sum to their respective prevalences:

\begin{equation}
    q_I(T) = \sum_{t = 0}^T f_I(t), \qquad
    q_V(T) = \sum_{t = 0}^T f_V(t),
    \label{eq:prev_f}
\end{equation}
and the prevalences are related:

\begin{equation}  q_N(T) + q_I(T) + q_V(T) = 1.
\label{eq:prev_relation}
\end{equation}
Motivated by the limiting behavior of antibody kinetics as in \cite{bedekar2022prevalence}, the na{\"i}ve distribution is identical to both the infected and vaccinated distributions on time step 0 of the personal timeline and asymptotically:

\begin{equation}
    N(\bm{r}) = R(\bm{r},0) = W(\bm{r},0) = \lim_{t \to \infty} R(\bm{r},t) = \lim_{t \to \infty} W(\bm{r},t).
    \label{eq:limiting}
\end{equation}
Replacement in Eq. \eqref{eq:Q} using Eq. \eqref{eq:P}, \eqref{eq:V}, and \eqref{eq:limiting} and combining and rearranging terms gives

\begin{equation}
  Q(\bm{r}, T) = N(\bm{r})  +  \sum_{t = 0}^{T-1}   \left[ R(\bm{r}, T-t) - N(\bm{r}) \right] f_I(t)  + \sum_{t = 0}^{T-1}  \left[ W(\bm{r}, T-t) - N(\bm{r}) \right] f_V(t) .
\label{eq:Q_orig}
\end{equation}

\subsection{Prevalence estimation}
\label{sec:prev_est}

Unbiased estimators can be constructed for the prevalences $q_N(T), q_I(T)$, and $q_V(T)$.  
Introduce a partition $\{D_1, D_2, D_3\}$ of the domain $\Omega$ such that
\begin{equation} 
D_1 \cup D_2 \cup D_3 = \Omega\ \text{ and } D_j \cap D_{\Tilde{j}} = \emptyset \ \ \forall j,\Tilde{j} \in \{1,2,3\} \text{ such that } j \neq \Tilde{j}.
\label{eq:partition}
\end{equation} 
Then define
\begin{equation}
\begin{split}
    Q_j(T) & = \int_{D_j} Q(\bm{r},T) d \bm{r}, \quad j = 1,2,3 \\
    & = N_j + \sum_{t = 0}^{T-1} \left[ R_j( T-t) - N_j \right] f_I(t)  +  \sum_{t = 0}^{T-1} \left[ W_j( T-t) - N_j \right] f_V(t),
    \label{eq:Q_j}
    \end{split}
\end{equation}
where

\begin{equation}
    N_j = \int_{D_j} N(r) d \bm{r}, 
    \label{eq:N_j}
\end{equation}

\begin{equation}
    R_j(T-t) = \int_{D_j} R(r, T-t) d \bm{r},
    \label{eq:R_j}
\end{equation}
and $W_j$ is defined similarly to $R_j$. Then, arbitrarily choosing to use $D_1$ and $D_2$, using Eq. \eqref{eq:Q_j}, for $T = 1$ we have

\begin{equation}
    Q_1(1) = N_1 + [R_1(1) - N_1]f_I(0) + [W_1(1) - N_1] f_V(0),
\end{equation}
and similarly

\begin{equation}
    Q_2(1) = N_2 + [R_2(1) - N_2]f_I(0) + [W_2(1) - N_2] f_V(0).
\end{equation}
In matrix form this is given by

\begin{equation}
    \begin{bmatrix}
        Q_1(1) \\
        Q_2(1)
    \end{bmatrix} = \begin{bmatrix}
        R_1(1) - N_1 & W_1(1) - N_1 \\
        R_2(1) - N_2 & W_2(1) - N_2
    \end{bmatrix}
    \begin{bmatrix}
        f_I(0) \\
        f_V(0)
    \end{bmatrix} + \begin{bmatrix}
        N_1 \\
        N_2
    \end{bmatrix}.
\end{equation}
To simplify notation, let $\bm{f}(0) = [f_I(0), f_V(0)]^T$, $\bm{M}(1) = \begin{bmatrix}
        R_1(1)  & W_1(1) \\
        R_2(1)  & W_2(1)
    \end{bmatrix}$, $\bm{N}^* = [N_1, N_2]^T [1,1]$, $\bm{N} = [N_1, N_2]^T$ and $\bm{Q}(1) = [Q_1(1), Q_2(1)]^T$. Then $\bm{Q}(1)$ can be written as
    
\begin{equation}
\bm{Q}(1) =  \left[\bm{M}(1) - \bm{N}^*  \right] \bm{f}(0) + \bm{N}.        
\end{equation}
Solving for $\bm{f}(0)$ gives

\begin{equation}
   \bm{f}(0)  = [\bm{M}(1) - \bm{N}^*]^{-1}
     [\bm{Q}(1) - \bm{N}].
\end{equation}
Define the Monte Carlo estimators $\hat{Q}_j(T)$ by

\begin{equation}
    Q_j(T) \approx \hat{Q}_j(T) = \frac{1}{S} \sum_{\ell = 1}^S \mathbb{I}_j(\bm{r}_{\ell}),
\end{equation}
where $\mathbb{I}_j$ is the indicator function on $D_j$ and $\{\bm{r}_1, \ldots, \bm{r}_S\}$ is the set of sample values observed on time step $T$ from   $S$ randomly-collected samples. Let $\bm{\hat{f}}(0) = [\hat{f}_I(0), \hat{f}_V(0)]^T$ and $\bm{\hat{Q}}(1) = [ \hat{Q}_1(1), \hat{Q}_2(1)]^T$. Then, we can estimate $\bm{f}(0)$, which contains the fractions of the population newly infected or vaccinated, on time step 0, via

    \begin{equation}
        \bm{f}(0) \approx \bm{\hat{f}}(0) = [\bm{M}(1) - \bm{N}^*]^{-1} [\bm{\hat{Q}}(1) - \bm{N}].
         \label{eq:f0_est}
    \end{equation}
 We will assume that the inverse in Eq. \eqref{eq:f0_est} exists; additional discussion follows in Section \ref{sec:network_approach}.

    We now iterate in this vein to obtain linear systems in terms of previously obtained estimates. Let $\bm{f}(t) = [f_I(t), f_V(t)]^T$, $\bm{M}(T-t) = \begin{bmatrix}
        R_1(T-t)  & W_1(T-t) \\
        R_2(T-t)  & W_2(T-t)
    \end{bmatrix}$,  and $\bm{\hat{Q}}(T) = [\hat{Q}_1(T), \hat{Q}_2(T)]^T$. We estimate
    the fraction of the population newly infected on time step $T-1$ to be
    
    \begin{equation}
        \hat{\bm{f}}(T-1) = [\bm{M}(1) - \bm{N}^*]^{-1} \left\{  \bm{\hat{Q}}(T) - \bm{N} - \sum_{t = 0}^{T-2} [\bm{M}(T-t) - \bm{N}^*] \bm{\hat{f}}(t) \right\}.
    \end{equation}
Notice that we use the Monte-Carlo estimate from the population at time step $T$ since the emergence of the disease to obtain information about the prevalence at the previous time step. This is because we cannot  immediately discern new infections or vaccinations from the na{\"i}ve population due to the time lag in personal antibody response \citep{bedekar2022prevalence}.
   The estimators $\bm{\hat{f}}(T-1)$ can be found recursively and then summed to estimate the prevalence at time step $T-1$:
    
    \begin{equation}
        \hat{\bm{q}}(T-1) = \sum_{t = 0}^{T-1} \hat{\bm{f}}(t),
    \end{equation}
    where the vector addition is computed element-wise and $\bm{\hat{q}}(T-1) = [\hat{q}_I(T-1), \hat{q}_V(T-1)]^T$. Then $\hat{q}_N(T-1) = 1 - \hat{q}_I(T-1) - \hat{q}_V(T-1)$. Note that it may be easiest to fix the partition $\{D_1, D_2, D_3\}$ to compute each $\hat{\bm{f}}(T-1)$ in the same manner.  See \ref{sec:unbiased} for a proof of the unbiasedness of the prevalence estimators, which follows from the fact that $\hat{Q}_j(\tau)$ is a Monte Carlo estimator of $Q_j(\tau)$.

    \section{Graph-theoretic approach}
    \label{sec:network_approach}

To begin this section, we reiterate that disallowing reinfection or revaccination significantly simplifies the task of finding $I(\bm{r}, T)$ and $V(\bm{r}, T)$ to a straightforward combination of our prior work \citep{bedekar2022prevalence,luke2023optimal}, as given by Eq. \eqref{eq:P} and \eqref{eq:V}. To then apply the method, one is left with a modeling exercise to construct  $R(\bm{r}, t)$ and $W(\bm{r},t)$ for the personal timeline antibody responses. However, if we allow reinfection and/or revaccination, the possible ways to arrive at an antibody level $\bm{r}$ at time step $T$ expand significantly, and a system that can track trajectories of infection and vaccination history becomes necessary. As an extreme example of the complexity, such a framework must be able to handle the situation in which at each next opportunity, an individual alternates between being infected and vaccinated, or repeats the same event over and over \citep{kocher2024adaptive}. Clearly, the corresponding models for $I(\bm{r}, T)$ and $V(\bm{r}, T)$ are not simple constructions, as they must take all possible trajectories--the collectively exhaustive set of events of interest--into account. This section revisits the simple problem of no reinfection, no revaccination through a different lens, with an eye towards the setting in which reinfection, revaccination, and movement between the two classes is allowed. 

 A generalization of the conditional probability models $I(\bm{r}, T)$ and $V(\bm{r}, T)$ from Section \ref{sec:prob_models} depends on  transitions into infection or vaccination states, as these affect antibody level. Thus, the models should depend on a weighted sum of all potential transitions, which can be represented via a transition matrix. The population-level antibody response over time can thus be formulated in terms of  the transition probabilities weighted by personal antibody response evolution. Given a current class and time $t$ in personal timeline, one can compute the transition probability for the next time step. This motivates using a Markov chain framework, because only the current state ($N, I, V$) and conditions $(\bm{r}, t, T)$ affect the next state. To consider the event of infection separately from previous infection, we partition class $I$ from Section \ref{sec:prob_models} into two states representing  new infections ($ \pos$) and previous infections ($\posp$). Similarly, we partition $V$ into $\vac$ and $\vacp$.

\subsection{Transition probabilities}

A transition matrix $S$ defines the probabilities of moving between states. Here,  $S(i,j)$ is the probability of moving \emph{to} state $i$ \emph{from} state $j$,  where the ordering is $N, \pos', \pos, \vac', \vac$. Let $T =0$ index the emergence of a disease. We assume that our initial state vector is $X_{-1} = \bm{e}_1$ to model the disease emergence, so that everyone is in state $N$ with probability 1 on the day before the disease emerges. Here, $\bm{e}_1$ is the first unit vector\footnote{This assumption holds for an emergent disease; we would  expect almost none of the population to be in the na{\"i}ve state in the case of an endemic virus such as the common cold.}. Let $X_j$ denote the state, or class, at time step $j$. We disallow transition from $\posp$ or $\vacp$ back to $N$, which is reasonable for a time scale on the order of several months. Denote the transition probabilities  by $s$. 
\begin{figure}[ht]
\centering
\includegraphics[scale=1.2]{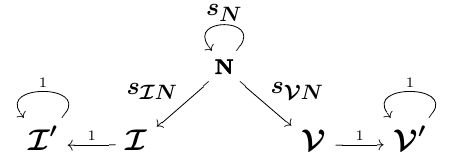}
\caption{Graph describing the allowable movements between states. Here, $N$ is na{\"i}ve, $\pos$ is newly infected, $\posp$ is previously infected,  $\vac$ is newly vaccinated, and $\vacp$ is previously vaccinated. Double subscripts on $s$ denote the transition probability from the second state to the first.}
        \label{fig:graph}
\end{figure}

We employ a graph to represent our framework, in which each state or class is a node and transitions between classes are directed, weighted edges; see Figure \ref{fig:graph}.
We let $s_N(T)$ denote the weight of the degenerate edge to $N$ from $N$, or the probability of staying na{\"i}ve. The probabilities $s_{\pos N}(T)$ and $s_{\vac N}(T)$ weight the edges to $\pos$ from $N$ and to $\vac$ from $N$, indicating infection or vaccination, respectively. Since we forbid reinfection or revaccination, one moves  to $\posp$ from $\pos$ or to $\vacp$ from $\vac$ on the next time step with probability 1. Once in $\posp$ or $\vacp$, one remains there with probability 1 and mounts their antibody response.

By definition, the transition probabilities depend on $f_I(T)$ and $f_V(T)$, the fractions of the population that are infected and   vaccinated on time step $T$, respectively. 
At time step $T$, the fraction of the population becoming newly infected or vaccinated is divided by the relative size of the na{\"i}ve population at the previous step to obtain the transition probability. This assumes that, of all the individuals in the na{\"i}ve class on the previous time step, a percentage $f_\pos$ or $f_\vac$ move to $\pos$ and $\vac$, respectively. Thus, we have 
     \begin{subequations}
        \begin{eqnarray}
            s_{\pos N}(T) = \frac{f_I(T)}{q_N(T-1)}, \\
            s_{\vac N}(T) = \frac{f_V(T)}{q_N(T-1)}, \\
            s_N(T) = 1 - s_{\pos N}(T) - s_{\vac N}(T).
        \end{eqnarray}
        \label{eq:sNs}
    \end{subequations}
\hspace{-3.5mm} The final relationship between the transition rates, the one defining $s_N(T)$, is a total probability statement. Here we define $q_N(-1) = 1$ to be consistent with our assumption that everyone is in the na{\"i}ve class on the day before the disease emerges. When $q_N(T-1) = 0$ we define $s_{\pos N}(T)$ and $s_{\vac N}(T)$ to be zero, as there are no individuals remaining in the $N$ class, so transitions out of that class are impossible. 
    
 We note a difference between the transition probabilities and the quantity $f(t)/q(T)$ in \cite{bedekar2022prevalence} as well as in Eq. \eqref{eq:P} and \eqref{eq:V}. In the context of these probabilistic models, this ratio $f(t)/q(T)$ is the proportion of those who became infected on time step $t$ of the disease emergence ($t<T$) out of the entire previously infected population as of time step $T$, and it is used to write out the total probability equation for time step $T$. In contrast, the transition probabilities $s_{\pos N}(T)$ and $s_{\vac N}(T)$ denote the fraction of the population that are newly infected or vaccinated \textit{out of those available}; i.e., the proportion of the population that was na{\"i}ve the time step before.
    
    The transition matrix for movement from time step $T-1$ to time step $T$ is thus given by
    
    \begin{equation}
    \label{eq:trans_mat_1step}
        S(T) = \begin{bmatrix}
            1 - s_{\pos N}(T) - s_{\vac N}(T) & 0 & 0 & 0 & 0 \\
            0 & 1 & 1 & 0 & 0 \\
            \frac{f_I(T)}{q_N(T-1)} & 0 & 0 & 0 & 0 \\
            0 & 0 & 0 & 1 & 1 \\
            \frac{f_V(T)}{q_N(T-1)} & 0 & 0 & 0 & 0 
        \end{bmatrix},
    \end{equation}
        where the ordering is $N, \pos', \pos, \vac', \vac$.
As an example, we distill the probability of a person ending up in a particular class after three time steps using product of transition matrices as follows: 

\begin{equation}
S(2)S(1)S(0)\mathbf{e}_1 = \begin{bmatrix}
s_N(2)s_N(1)s_N(0) \\
s_{\pos N}(0) + s_{\pos N}(1) s_N(0) \\
s_{\pos N}(2) s_N(1) s_N(0) \\
s_{\vac N}(0) + s_{\vac N}(1) s_N(0) \\
s_{\vac N}(2) s_N(1) s_N(0)
\end{bmatrix}.
    \end{equation}
The interpretation of the first three entries of the resulting vector is as follows: the probability that one stays na{\"i}ve through time step 3; the probability that one got infected on time step 1 and is considered previously infected on time step 2 plus the probability that one got infected on time step 2 and is considered previously infected on time step 3; and the probability that one did not get infected on time steps 1 or 2 but did get infected on time step 3.  In short, we confirm that all possible trajectories to $N$, $\posp$, and $\pos$ on time step 3 are considered. Parallel interpretations hold for $\vacp$ and $\vac$ as for $\posp$ and $\pos$. 

    Specifically, such multi-step transitions  can be represented in terms of the entries of the matrix-vector product of the subsequent transitions. A $\tau$ step transition from time step $0$ to time step $\tau$ is represented by $H_\tau,$
    
    \begin{equation} 
    H_\tau = S(\tau) S(\tau-1)\cdots S(1)S(0) = \left(\prod_{t = 0}^{\tau} S(\tau - t) \right), \end{equation}
    where $\tau - t$ is used to ensure indexing of the product in the correct order. Thus, if everyone starts in the na{\"i}ve class on time step $0$, then the total probabilistic distribution of the classes after $T$ time steps in the absolute timeline will be 

\begin{equation}
\begin{bmatrix}
\text{Prob}(X_{T} = N)  \\
\text{Prob}(X_{T} = \posp)\\
\text{Prob}(X_{T} = \pos)  \\
\text{Prob}(X_{T} = \vacp) \\
\text{Prob}(X_{T} = \vac)
\end{bmatrix} = H_T \bm{e}_1.
\label{eq:probH_vector}
\end{equation}

\subsection{Equivalence of state probabilities and original framework prevalences}

Here, we continue to assume that everyone begins in the na{\"i}ve class at time step zero, or that our initial state vector is $\bm{e}_1$.
By using Eq. \eqref{eq:probH_vector}, the state probabilities are given explicitly by

\begin{subequations}
\begin{align}
\text{Prob}(X_T = N)  & = \prod_{t = 0}^{T} s_N(t), \\
  \text{Prob}(X_T = \posp) &  = \sum_{t = 0}^{T-1} s_{\pos N}(t) \prod_{\tau = 0}^{t-1} s_N(\tau), \label{eq:state_P'}\\
 \text{Prob}(X_T = \pos) & = s_{\pos N}(T) \prod_{t = 0}^{T-1} s_N(t), \label{eq:state_P}\\
    \text{Prob}(X_T = \vacp) & = \sum_{t = 0}^{T-1} s_{\vac N}(t) \prod_{\tau = 0}^{t-1} s_N(\tau), \\
   \text{Prob}(X_T = \vac) & = s_{\vac N}(T) \prod_{t = 0}^{T-1} s_N(t).
\end{align}
\label{eq:state}
\end{subequations}
\hspace{-3.5mm} The state probabilities and the prevalences in Section \ref{sec:prev_est} are related; in fact, $ {\rm Prob} (X_T = N) = q_N(T)  , {\rm Prob} (X_T = \posp) = q_I(T-1), {\rm Prob} (X_T = \pos) = f_I(T),$ ${\rm Prob} (X_T = \vacp) = q_V(T-1),$ and $ {\rm Prob} (X_T = \vac) = f_V(T)$, as proved below. Using the definition of the state probability and Eq. \eqref{eq:sNs}, we have

\begin{align}
    {\rm Prob} (X_{T} = N)   = \prod_{t = 0}^{T} s_N(t)  & = \prod_{t = 0}^{T} \left[ 1 - s_{\pos N}(t) - s_{\vac N}(t) \right] = \prod_{t = 0}^{T} \left[ 1 - \frac{f_I(t)}{q_N(t-1)} - \frac{f_V(t)}{q_N(t-1)} \right].
    \end{align}
Using Eqs. \eqref{eq:prev_f} and  \eqref{eq:prev_relation} and the fact that $q_N(0) = 1$, we then find
    
\begin{equation}
\begin{split}
   {\rm Prob} (X_{T} = N)  
    & = \prod_{t = 0}^{T} \left[  \frac{q_N(t-1) - f_I(t) - f_V(t)}{q_N(t-1)} \right] \\
    & =  \prod_{t = 0}^{T} \left[  \frac{1 - q_I(t-1) - q_V(t-1) - f_I(t) - f_V(t)}{q_N(t-1)} \right] \\
    & = \prod_{t = 0}^{T} \left[  \frac{1 - q_I(t) - q_V(t)}{q_N(t-1)} \right] \\
    & = \prod_{t = 0}^{T}   \frac{ q_N(t)}{q_N(t-1)}  
    = q_N(T).
    \end{split}
    \label{eq:N_state}
\end{equation}
Using Eq. \eqref{eq:N_state}, we find that for $\posp$,

\begin{equation}
    {\rm Prob} (X_{T} = \posp)   = \sum_{t = 0}^{T-1} s_{\pos N} (t) \prod_{\tau = 0}^{t-1} s_N(\tau)   = \sum_{t = 0}^{T-1} \frac{f_I(t)}{q_N(t-1)} q_N(t-1) = \sum_{t = 0}^{T-1} f_I(t)  = q_I(T-1).
    \label{eq:prev_P'}
    \end{equation}
    This is expected, because the previously infected class does not include new infections occurring on time step $T$. Finally, using Eq. \eqref{eq:N_state}, for $\pos$ we find

\begin{equation}
{\rm Prob}(X_T = \pos) = s_{\pos N}(T) \prod_{t = 0}^{T-1} s_N(t) = s_{\pos N}(T) q_N(T-1) = \frac{f_I(T)}{q_N(T-1)} q_N(T-1) = f_I(T).
\label{eq:prev_P}
\end{equation}
One can analogously show that ${\rm Prob} (X_T = \vacp) = q_V(T-1)$ and ${\rm Prob} (X_T = \vac) = f_V(T)$. Thus,

\begin{equation}
\begin{split}
  {\rm Prob} (X_T = N) + {\rm Prob} (X_T = \posp) + {\rm Prob} (X_T = \pos) + {\rm Prob} (X_T = \vacp) + {\rm Prob} (X_T = \vac) \\
  = q_N(T) + q_I(T-1) + f_I(T) + q_V(T-1) + f_V(T) = q_N(T) + q_I(T) + q_V(T) = 1. 
  \end{split}
\end{equation}
 Since we have split the infected class of Section \ref{sec:methods} into $\pos$ and $\posp$ to represent new and old infections, we expect that ${\rm Prob} (X_T = \posp) + {\rm Prob} (X_T = \pos) = q_I(T)$; this is in fact true, as $q_I(T-1) + f_I(T) = q_I(T)$. This allows us to confirm the relationship between the state probabilities and the prevalences of Section \ref{sec:prev_est} and that the state probabilities sum to 1, as expected.

Eqs. \eqref{eq:N_state}, \eqref{eq:prev_P'}, \eqref{eq:prev_P} and analogous statements for $\vacp, \vac$ help  rewrite the $\tau-$ step transition matrix $H_\tau$ as follows:

\begin{equation}
    H_{\tau} = \begin{bmatrix}
    q_N(\tau) & 0 & 0 & 0 & 0\\
    \sum\limits_{t=0}^{\tau-1} f_I(t) & 1 & 1 & 0 & 0\\
    f_I(\tau) & 0 & 0 & 0 & 0\\
    \sum\limits_{t=0}^{\tau-1} f_V(t) & 0 & 0 & 1 & 1\\
    f_V(\tau) & 0 & 0 & 0 & 0
\end{bmatrix},
\label{eq:H}
\end{equation}

where the ordering is $N, \pos', \pos, \vac', \vac$. Notice that these transitions are as expected. However, these transition matrices do not carry over information about antibody kinetics, nor the convolution between personal and absolute timelines that leads to a given sampled antibody value from a population on a given time. We will incorporate this in the next subsection.

\subsection{Conditional probabilities in terms of transition matrices and personal timeline models}

The set of previously infected individuals can be partitioned using the time period when they were newly infected, i.e., when they were in the transient class $\posp$. Thus, using the law of total probability, the conditional probability density  for an antibody measurement $\bm{r}$ during time step $T$ in the absolute timeline given that the sample comes from a previously infected individual is 

\begin{equation}
  \hspace{-8mm} \text{Prob}(\bm{r},T | X_T = \posp) = \sum\limits_{t=0}^{T-1}  \text{Prob}(\bm{r},T, X_{t} = \pos | X_T = \posp) = \frac{1}{\text{Prob}(X_T = \posp)}\sum\limits_{t=0}^{T-1}  \text{Prob}(\bm{r},T, X_{t} = \pos, X_T = \posp).
\end{equation}
This summand consists of people who were na{\"i}ve through time period $t-1$, become newly infected in time period $t$, and stay in the previously infected class thereafter, i.e. 
$NN\cdots N \pos \posp \posp \cdots \posp$.  
For such a sequence, $R(\bm{r},T-t)$ is the distribution of the antibody response on time step $T$ in the absolute timeline. Thus,

\begin{equation}
\hspace{-10mm}    \text{Prob}(\bm{r},T, X_{t} = \pos, X_T = \posp) = R(\bm{r},T - t) s_{\pos N}(t) \prod_{\tau = 0}^{t-1} s_N(\tau) = R(\bm{r},T - t) \left<\left(\prod_{\tau = 0}^{t} S(t-\tau) \right) \bm{e}_1, \bm{e}_3\right>.
\end{equation}
Here, angle brackets denote the (dot) inner product.
In total, the conditional probability density can be rewritten as

\begin{equation}
\begin{split}
 \text{Prob}(\bm{r},T | X_T = \posp) & = \frac{1}{\text{Prob}(X_T = \posp)} \left(\sum_{t = 0}^{T-1} R(r,T - t) \left<\left( \prod_{\tau = 0}^{t} S(t-\tau) \right) \bm{e}_1, \bm{e}_3\right> \right)\\
 & = \frac{1}{\left<H_T \bm{e}_1, \bm{e}_2\right>} \left(\sum_{t = 0}^{T-1} R(r,T - t) \left<H_t \bm{e}_1, \bm{e}_3\right> \right).\\
\end{split}
\label{eq:cond_prob_P'}
\end{equation}
In other words, the inner product inside the large parentheses is the prevalence of newly infected individuals on a particular time step. Thus, the sum in the last term is the inner product of responses on different time steps with the vector of newly infected prevalences.
Using Eq. \eqref{eq:state_P'}-\eqref{eq:state_P}, we can see that this is solely in terms of the transition matrix $S$ and the personal antibody response model $R$. The expression for ${\rm Prob} (\bm{r}, T | X_T = \vacp)$ is analogous.

The expression for ${\rm Prob} (\bm{r}, T | X_T = \pos)$ is simpler than that for $\posp$ because there is only one possible sequence of state transitions: $N N \cdots N \pos$, where the transition from $N$ to $\pos$ occurs on time step $T$. This sequence has antibody response distribution $R(\bm{r}, 0 ) = N(\bm{r})$, and thus ${\rm Prob} (\bm{r}, T | X_T = \pos)  = {\rm Prob} (\bm{r}, T | X_{T-1} = N, X_T = \pos) = N(\bm{r})$. 

\subsubsection{Equivalence of measurement density in graph-theoretic and original frameworks}
\label{sec:equiv_meas_dens}

Ideas similar to Eq. \eqref{eq:Q} lead us to derive the measurement density,

\begin{align}
\hspace{-5mm}\begin{split}
    Q(\bm{r}, T) & = q_N(T) {\rm Prob} (\bm{r}, T | X_T = N) + q_{\posp}(T) {\rm Prob} (\bm{r}, T | X_T = \posp) + q_{\pos}(T) {\rm Prob} (\bm{r}, T | X_T = \pos) \\
    & + q_{\vacp}(T) {\rm Prob} (\bm{r}, T | X_T = \vacp) + q_{\vac}(T) {\rm Prob} (\bm{r}, T | X_T = \vac).
    \end{split}
    \label{eq:Q_network}
\end{align}
Using Eqs. \eqref{eq:state} and \eqref{eq:cond_prob_P'}, the relationship between the state probabilities and the prevalences in Section \ref{sec:prev_est}, and the above discussion, after some rearranging, Eq. \eqref{eq:Q_network} can be written as 

\begin{align}
\begin{split}
    Q(\bm{r},T) & =  q_N(T-1) N(\bm{r}) + \sum_{t = 0}^{T-1} R(\bm{r},T - t) \left<\left( \prod_{\tau = 0}^{t} S(t-\tau) \right) \bm{e}_1, \bm{e}_3\right>  \\
    &  + \sum_{t = 0}^{T-1} W(\bm{r},T - t) \left<\left( \prod_{\tau = 0}^{t} S(t-\tau) \right) \bm{e}_1, \bm{e}_5\right> .
    \end{split}
    \label{eq:Q_graph}
\end{align}
Here we have also used $q_N(T-1) = q_N(T) + f_I(T) + f_V(T)$.
 While Eq. \eqref{eq:Q_graph} is a different representation of the measurement density from that given by Eq. \eqref{eq:Q_orig} in Section \ref{sec:methods}, by rewriting the inner products as state probabilities and using the relationship between the state probabilities and the prevalences in Section \ref{sec:prev_est}, one can show their equivalence: 
 
\begin{equation}
\hspace{-5mm}\begin{split}
    Q(\bm{r},T) & = \left[ 1 - q_I(T-1) - q_V(T-1) \right] N(\bm{r}) 
      + \sum_{t = 0}^{T-1} R(\bm{r},T - t) \text{Prob}(X_t = \pos)    + \sum_{t = 0}^{T-1} W(\bm{r},T - t) \text{Prob}(X_t = \vac) \\
    & = \left[ 1 - \sum_{t = 0}^{T-1} f_I(t) - \sum_{t = 0}^{T-1} f_V(t) \right] N(\bm{r})  + \sum_{t = 0}^{T-1} R(\bm{r},T - t) f_I(t)   + \sum_{t = 0}^{T-1} W(\bm{r},T - t) f_V(t)  \\
    & = N(\bm{r}) + \sum_{t = 0}^{T-1}  [R(\bm{r},T - t) - N(\bm{r}) ]f_I(t)   + \sum_{t = 0}^{T-1} [W(\bm{r},T - t) - N(\bm{r})] f_V(t) .
    \end{split}
\end{equation}
Due to this equivalence, prevalence estimation in this graph-theoretic approach will follow Section \ref{sec:prev_est}. Then, original class $I$ will be broken into $\pos$ and $\posp$ so that the estimators are $\hat{q}_{\posp}(T) = \sum_{t = 0}^{T-1} \hat{f}_I(t)$ and $\hat{q}_{\pos}(T) = \hat{f}_I(T)$ by their definitions. Similar estimators can be found for $q_{\vacp}(T)$ and $q_{\vac}(T)$.

\subsection{Estimation of transition probability matrices}
\label{sec:tran_est}

Let us reconsider
the measurement density  Eq. \eqref{eq:Q_graph}, from Section \ref{sec:equiv_meas_dens}. We note that it can be written in terms of the transition matrix $H_t$ from Eq. \eqref{eq:H} as 
\begin{align}
\begin{split}
    Q(\bm{r},T) & = q_N(T-1) N(\bm{r}) + \sum_{t = 0}^{T-1} R(\bm{r},T - t) H_{t,(3,1)}   + \sum_{t = 0}^{T-1} W(\bm{r},T - t) H_{t,(5,1)},
    \end{split}
\end{align}
where $H_{t,(k,1)} := \left< H_t \bm{e}_1, \bm{e}_k\right>.$ Integrating both sides of the equation over some subdomain $D_j$ of the antibody measurement space leads to 
\begin{align}
\begin{split}
    Q_{D_j}(T) & = q_N(T-1) N_{D_j} + \sum_{t = 0}^{T-1} R_{D_j}(T - t) H_{t,(3,1)}    + \sum_{t = 0}^{T-1} W_{D_j}(T - t) H_{t,(5,1)}.
    \end{split}
    \label{eq:totalProb}
\end{align}

Unbiased estimation of $Q_{D_j}$ can be achieved by Monte-Carlo estimation as in Section \ref{sec:prev_est}: by measuring antibody levels for randomly selected samples from the population during time period $T$, followed by a counting of the fraction of the measurements that fall in that subdomain.

Notice that for $T = 0,$ we have $Q_{D_j}(0) = N_{D_j}$; this follows our assumption that every person starts  na{\"i}ve at $T = -1$, the day before the disease emerges, and that antibody response does not mount immediately after infection. That is, the total probability mass is distributed exactly as the na{\"i}ve distribution: this is as expected. For $T=1$,
\begin{equation}   
Q_{D_j}(1)  = q_N(0)  N_{D_j} + R_{D_j}(1) H_{0,(3,1)}  +  W_{D_j}(1) H_{0,(5,1)}.
\label{eq:totalProbT1}
\end{equation}

Let $D_1, D_2, D_3$ partition the domain as in Eq. \eqref{eq:partition}.
Notice that for such a partition, as $N, R(t), W(t)$ are probability distributions for all $t\geq0$,
\begin{equation}
\hspace{-5mm} N_{D_1} + N_{D_2} + N_{D_3} = R_{D_1}(t) + R_{D_2}(t) + R_{D_3}(t) = W_{D_1}(t) + W_{D_2}(t) + W_{D_3}(t) = 1 = Q_{D_1}(t) + Q_{D_2}(t) + Q_{D_3}(t).
\label{eq:totalTo1}
\end{equation}
Using Eq. \eqref{eq:totalProbT1} for $D_1, D_2, D_3$, we can write down a system of linear equations, given by
\begin{equation}
\label{eq:tranProbEst0}
    \begin{bmatrix}
N_{D_1} & R_{D_1}(1) & W_{D_1}(1)\\
N_{D_2} & R_{D_2}(1) & W_{D_2}(1)\\
        1 & 1 & 1
    \end{bmatrix} \begin{bmatrix}
        q_N(0) \\ H_{0,(3,1)}\\ H_{0,(5,1)}
    \end{bmatrix} = \begin{bmatrix}
        Q_{D_1}(1) \\ Q_{D_2}(1)\\ q_N(-1)
    \end{bmatrix} \approx \begin{bmatrix}
        \widehat{Q}_{D_1}(1) \\ \widehat{Q}_{D_2}(1)\\ q_N(-1)
    \end{bmatrix}
\end{equation}
where the term $q_N(-1)=1$ as before due to our assumption that everyone starts na{\"i}ve before the emergence of the disease. The last equation of this matrix system arises out of an application of Eq. \eqref{eq:totalTo1}, and expresses that the na{\"i}ve population in the preceding time step distributes into na{\"i}ve, newly infected, and newly vaccinated in the next time step. 

We can thus estimate $q_N(0), H_{0,(3,1)}, H_{0,(5,1)}$ and obtain the respective estimates $\widehat{q_N}(0), \widehat{H}_{0,(3,1)},$ and $ \widehat{H}_{0,(5,1)}$. 
Via induction, for a general $T$ we obtain the following system
\begin{multline}
\label{eq:tranProbEstT}
        \begin{bmatrix}
        N_{D_1} & R_{D_1}(1) & W_{D_1}(1)\\
        N_{D_2} & R_{D_2}(1) & W_{D_2}(1)\\
        1 & 1 & 1
    \end{bmatrix} \begin{bmatrix}
        q_N(T-1) \\ H_{(T-1)(3,1)}\\ H_{(T-1)(5,1)}
    \end{bmatrix}\\
    = \begin{bmatrix}
        Q_{D_1}(T) - \sum\limits_{t=0}^{T-2} \left( R_{D_1}(T-t) H_{t,(3,1)} + W_{D_1}(T-t) H_{t,(5,1)} \right)\\ Q_{D_2}(T) - \sum\limits_{t=0}^{T-2} \left( R_{D_2}(T-t) H_{t,(3,1)} + W_{D_2}(T-t) H_{t,(5,1)} \right)\\ q_N(T-2)
    \end{bmatrix}\\
    \approx \begin{bmatrix}
    \widehat{Q}_{D_1}(T) - \sum\limits_{t=0}^{T-2} \left( R_{D_1}(T-t) \widehat{H}_{t,(3,1)} + W_{D_1}(T-t) \widehat{H}_{t,(5,1)} \right)\\ \widehat{Q}_{D_2}(T) - \sum\limits_{t=0}^{T-2} \left( R_{D_2}(T-t) \widehat{H}_{t,(3,1)} + W_{D_2}(T-t) \widehat{H}_{t,(5,1)} \right)\\ \widehat{q}_N(T-2)
    \end{bmatrix}.
\end{multline}
Note that even though we only estimate values pertaining to the na{\"i}ve, newly infected, and newly vaccinated entry in the first column, these sequential estimates determine the values for previously infected and vaccinated entries by the following recursive relation:

\begin{equation}
\label{eq:tranProbH2}
\begin{split}
    H_{(T-1)(2,1)} & = H_{(T-2)(2,1)} + H_{(T-2)(3,1)}\\
    & = H_{(T-3)(2,1)} + H_{(T-3)(3,1)} + H_{(T-2)(3,1)} = \cdots \\
    & = H_{0(2,1)} + \sum\limits_{\tau=0}^{T-2} H_{\tau(3,1)}\\
    & = \sum\limits_{\tau=0}^{T-2} H_{\tau(3,1)}  \approx \sum\limits_{\tau=0}^{T-2} \widehat{H}_{\tau(3,1)}.
    \end{split}
\end{equation}

and similarly

\begin{equation}
\label{eq:tranProbH4}
H_{(T-1)(4,1)} = H_{(T-2)(4,1)} + H_{(T-2)(5,1)} = ... = H_{0(4,1)}+\sum\limits_{\tau=0}^{T-2} H_{\tau(5,1)} = \sum\limits_{\tau=0}^{T-2} H_{\tau(5,1)}  \approx \sum\limits_{\tau=0}^{T-2} \widehat{H}_{\tau(5,1)}.
\end{equation}
This is as expected because the prevalence of previously infected/vaccinated is obtained as an accumulation of prevalences for newly infected/vaccinated over the entire absolute timeline.

\section{Example applied to SARS-CoV-2 antibody data}
\label{sec:example}

\begin{figure}[ht]
    \centering
\includegraphics[scale = 0.45]{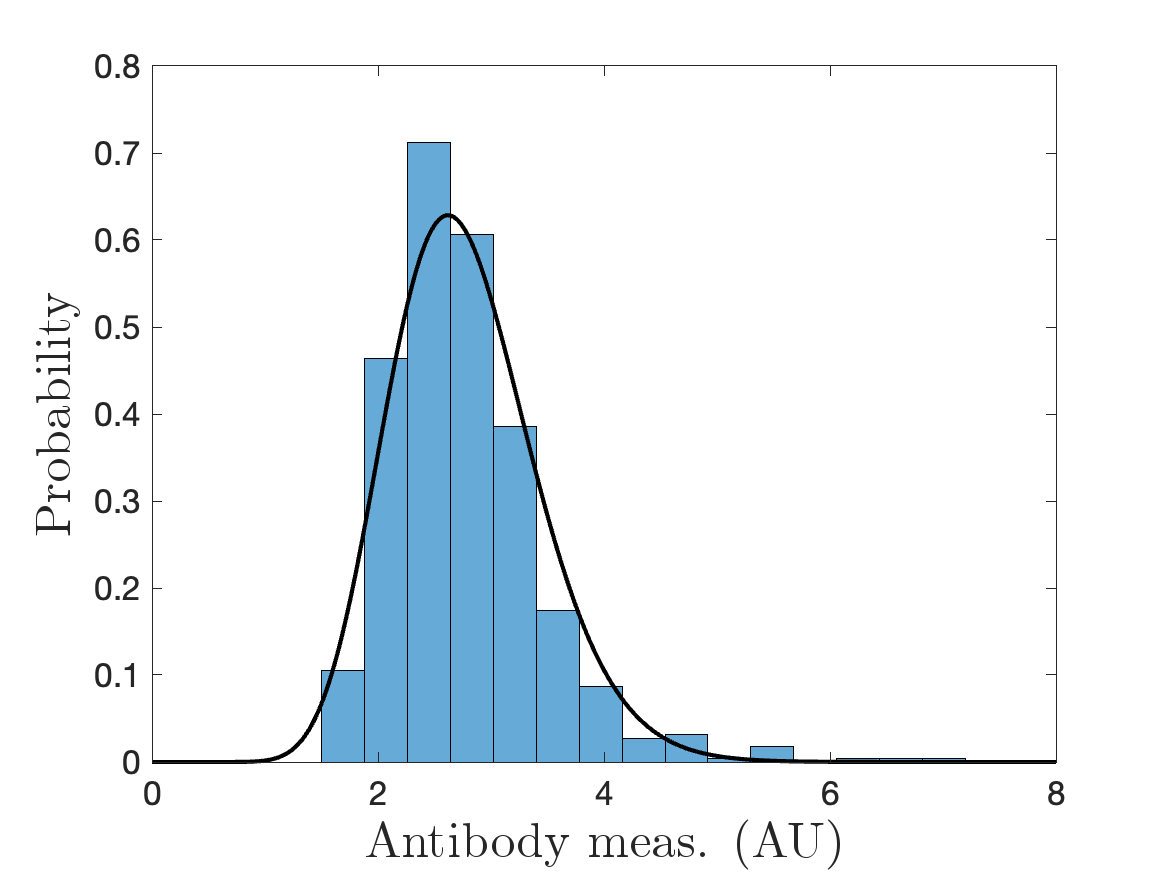}
\caption{Log-transformed synthetic antibody measurements from the na{\"i}ve population with corresponding probability model.}
\label{fig:N_model}
\end{figure}

\begin{figure}[ht]
\centering
\includegraphics[scale = 0.43]{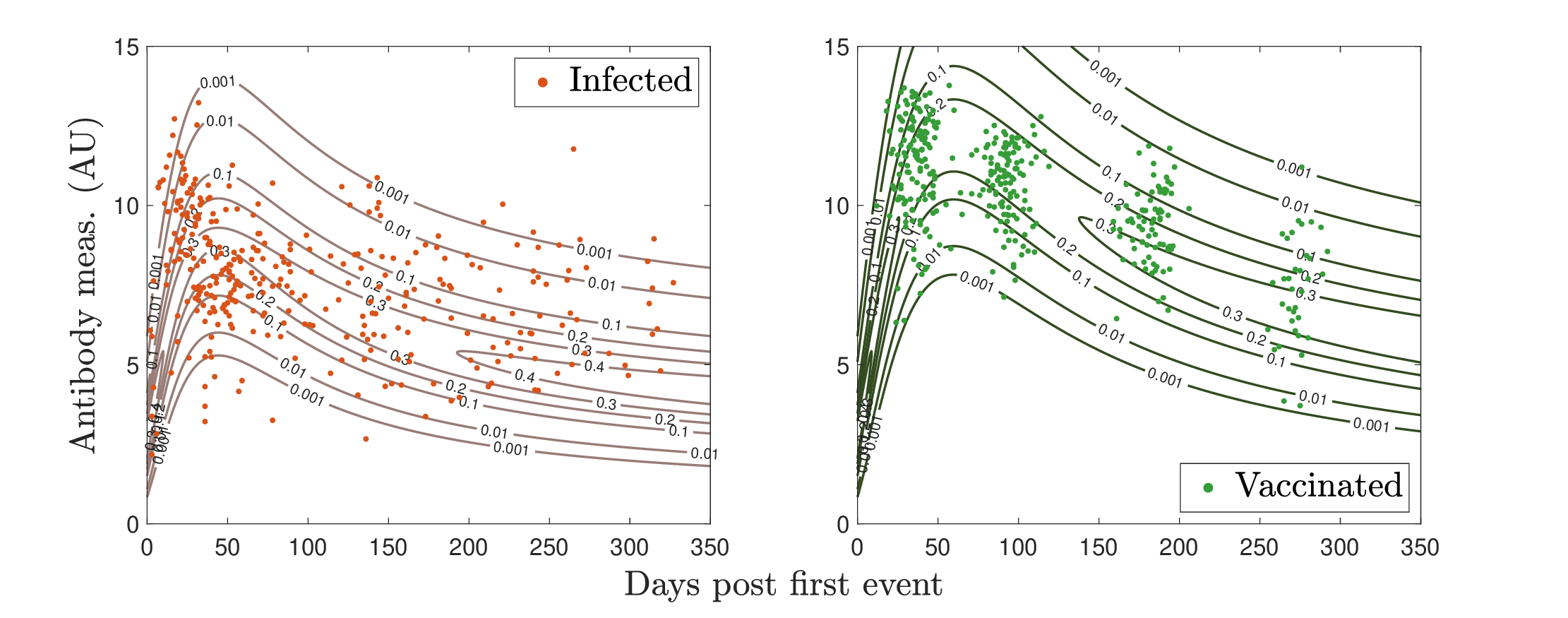}
\caption{Log-transformed synthetic antibody measurements from the infected and vaccinated populations with corresponding probability models.}
\label{fig:PV_models}
\end{figure}

In the context of SARS-CoV-2, our models characterize the time frame around spring 2021, when individuals either had a previous infection or were receiving their first vaccination, with very few people having done both.
We create synthetic training data motivated by clinical data from \cite{abela2021multifactorial} and  \cite{congrave2022twelve}. 
The synthetic data are created by studying immunoglobulin G (IgG) measurements for na{\"i}ve,  infected, and vaccinated individuals. These values are considered together to have arbitrary units (AU) and log-transformed similarly to \cite{patrone2021classification}, \cite{bedekar2022prevalence}, and \cite{luke2023optimal} to yield the unit-less, one-dimensional measurement 

\begin{equation}
    r = \log_2(\tilde{r}).
\end{equation} 
 We use gamma distributions to model the  infected and vaccinated antibody responses $t$ days after infection or vaccination, which both change with time, and the na{\"i}ve distribution. We use pre-vaccine measurements reported as SARS-CoV-2-na{\"i}ve to model the na{\"i}ve population with
 
\begin{equation}
    N(r) = \frac{1}{\Gamma(\alpha_n) \beta_n^{a_n}} r^{\alpha_n-1} e^{-r/\beta_n}.
\end{equation}
For this synthetic dataset, $\alpha_n = 15.1$ and $\beta_n = 0.184$.
We allow $\alpha$ to vary in time for both the vaccinated and infected responses as

\begin{equation}
    \alpha_c(t) = \frac{\theta_{1,c} t}{1 + \theta_{2,c} t^2} + \alpha_n, \text{ where } c \in \{i,v\},
\end{equation}
 where the subscripts $i$ and $v$ denote  infected and vaccinated. The model for the personal timeline of an infected individual is then given by

\begin{equation}
    R(r,t) = \frac{1}{\Gamma(\alpha(t)) \beta_n^{a(t)}} r^{\alpha(t)-1} e^{-r/\beta_n}.
    \label{eq:vax_model}
\end{equation}
Following \cite{bedekar2022prevalence}, we require that at $t = 0$, both the vaccinated and infected models are identical to $N(r)$. This is enforced by our modeling; note that the shape and scale at $t=0$ and as $t\rightarrow\infty$ are identical to those for na{\"i}ve. 
The model for personal timeline for  vaccination, $W(r,t)$, is given similarly to Eq. \eqref{eq:vax_model}.  The model parameters are $\theta_{1, i} = 1.56, \theta_{2, i} = 5.1 \times 10^{-4}, \theta_{1, v} = 1.74$, and $\theta_{2, v} = 2.8 \times 10^{-4}$.
 The models are shown in Figures \ref{fig:N_model} and \ref{fig:PV_models}.

\subsection{Prevalence estimation via transition probability matrices}

We conduct prevalence estimation for test data sampled from the probability models described above and through transition probability matrix estimation methods developed in  Section \ref{sec:tran_est}. 
We create 1000 sets of synthetic test data for various numbers of sample points $N_s$ to calculate the mean and standard deviation of the prevalence estimates. We discretize time into time steps of $dt = 21$ days and use 10 time periods. This results in 210 days total, or about 7 months, which is on the order of the synthetic training data.

To mimic a wave of infections during the emergence of a disease, we assume a sinusoidal change in the  infected prevalence per time period, given by

 \begin{equation}
f_I(0) = 0.01, \quad f_I(t) = 0.01 \sin \left( \frac{\pi t}{10} \right), \quad t \in \{1, \ldots, 10\}.
 \end{equation}
 This gives the corresponding true  newly plus previously infected prevalence as 
 
 \begin{equation}
q_I(t) = 
0.01 \left[ 1 + \sum_{\tau = 0}^t \sin \left( \frac{ \pi \tau }{10} \right) \right], \quad t \in \{0, 1, \ldots, 10\}. 
 \end{equation}
 We also assume a constant rate of vaccination, given by an incidence
 
 \begin{equation}
     f_V(t) = 0.01, \quad t \in \{0, 1, \ldots, 10\},
 \end{equation}
 and corresponding true newly and previously vaccinated prevalence as
 
 \begin{equation}
     q_V(t) = 0.01 t  + 0.01, \quad t \in \{0, 1, \ldots, 10\}.
 \end{equation}
Thus, the prevalence of na{\"i}ve for time-period $t$ is 

\begin{equation}
    q_N(t) = 1 - q_V(t) - q_I(t) = 0.08 -  0.01 t -0.01 \sum_{\tau = 0}^t \sin \left( \frac{ \pi \tau }{10} \right), \quad t \in \{0, 1, \ldots, 10\}. 
\end{equation}

These incidence rates lead to the true one-step and $\tau$-step transition matrices by using Eq. \eqref{eq:trans_mat_1step} and Eq. \eqref{eq:H} respectively. The $\tau$-step transition matrix entries are then estimated using methods detailed in Section \ref{sec:tran_est}, which provide us with estimates of prevalence for different classes. 

\begin{figure}[ht]
\centering
\subfloat[][Wave for previously infected]{\includegraphics[scale=0.34]{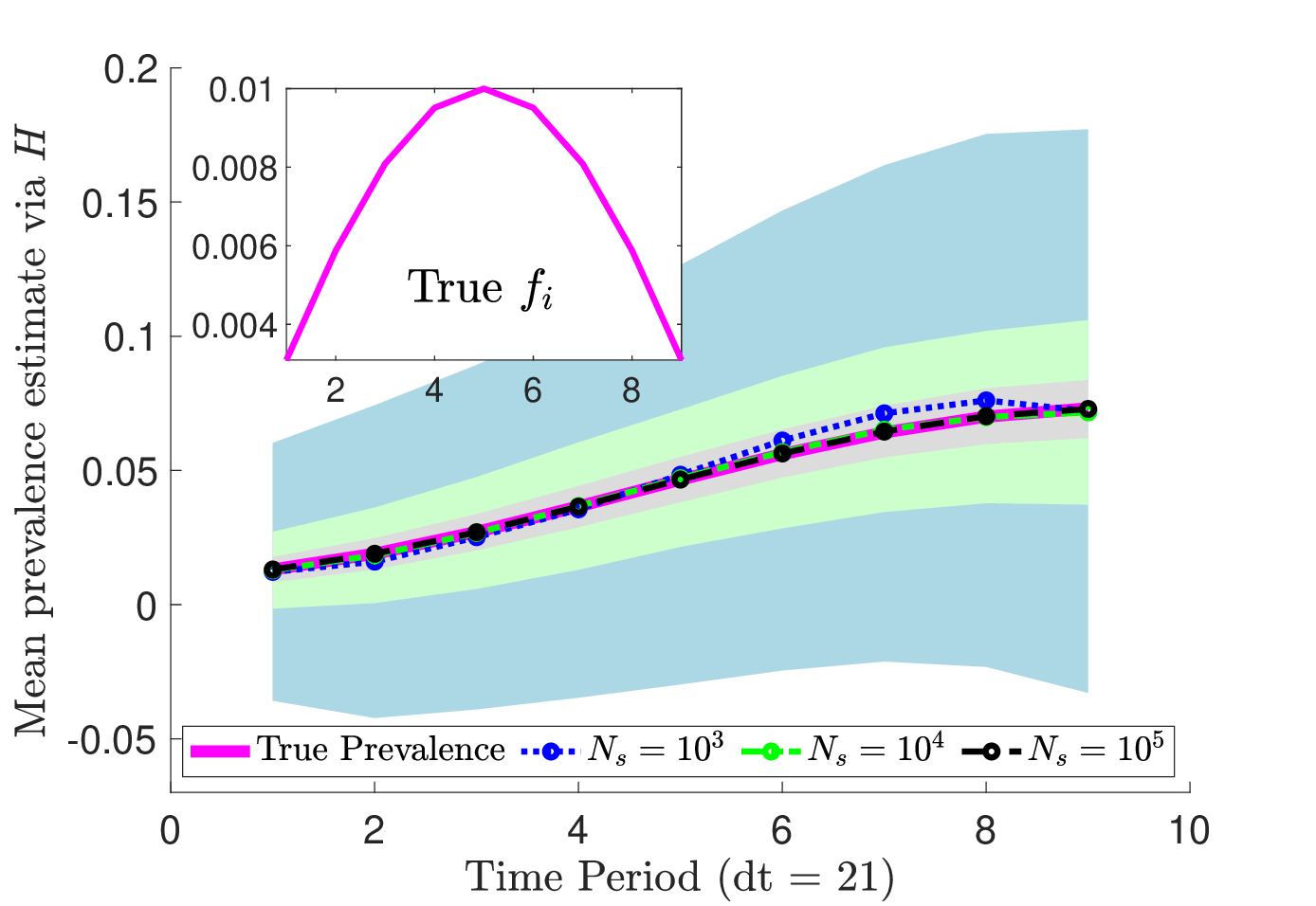}}
\subfloat[][Constant newly vaccinated]{\includegraphics[scale=0.34]{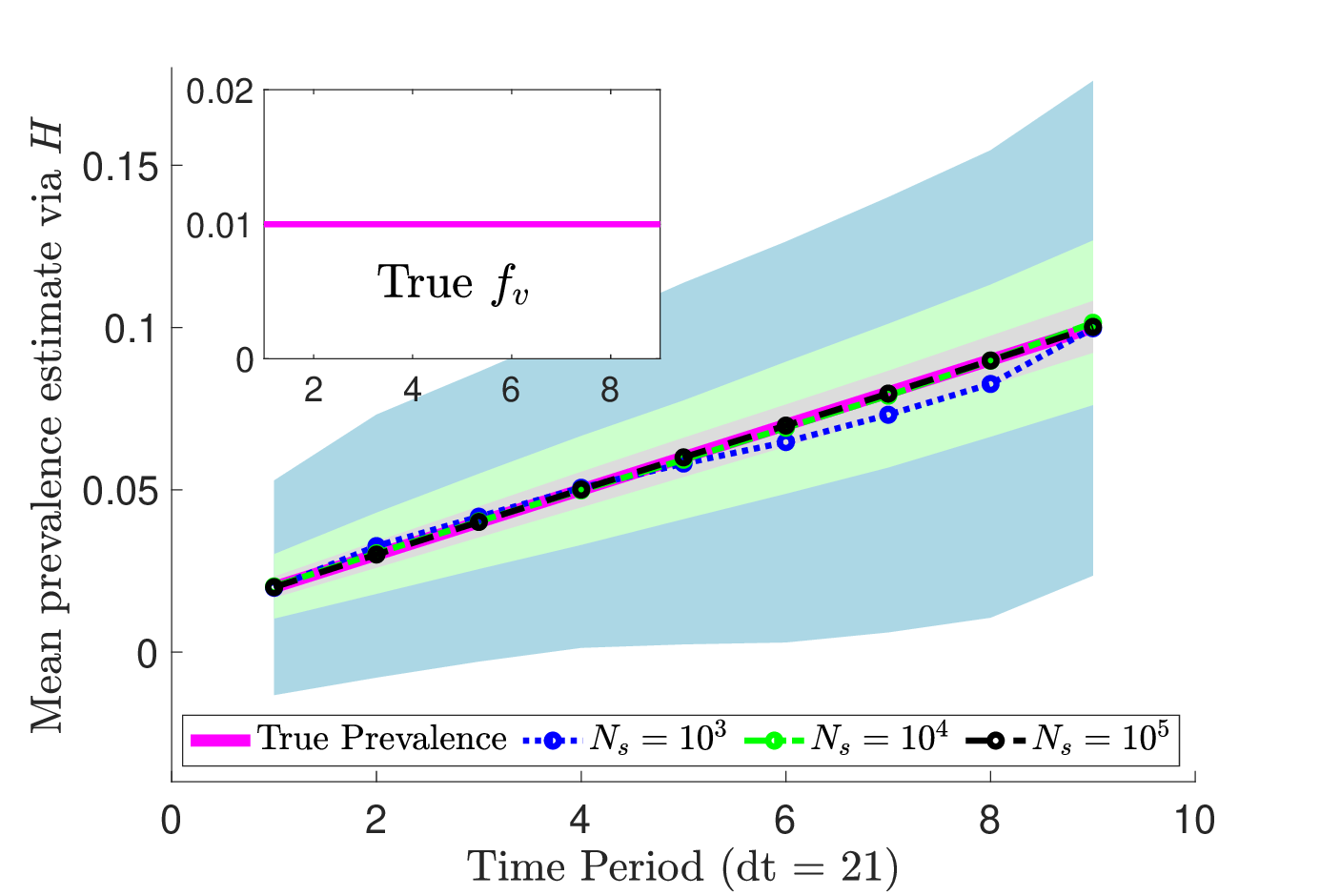}}
    \caption{Prevalence estimation via transition probability estimates using synthetic data for antibody responses. The mean over 1000 synthetic data sets is shown for various numbers of samples $N_s$ with standard deviation confidence intervals (shown in a lighter shade of the corresponding color of $N_s$) over time.}
    \label{fig:prev_est_H_not_sep}
\end{figure}

 The results of prevalence estimation are shown in Figure \ref{fig:prev_est_H_not_sep}. The mean prevalence estimates are shown as data points and corresponding standard deviations are shown as shaded regions in lighter colors. The mean estimates agree fairly well across the sample sizes and with the true values in both the  infected and vaccinated cases. We note that for low sample sizes, such as $N_s = 10^3$, some prevalence estimates are negative, which are infeasible. However, the standard deviation of the estimates decreases with increasing sample size as expected. We observe larger prevalence estimation standard deviations at later time periods, as expected following  \cite{bedekar2022prevalence}, who noted that errors accumulate over time. For $N_s = 10^5$, the average prevalence estimate errors across all time periods are $(17 \pm 15)$ \% for infected and $(8.6 \pm 7.0)$ \% for vaccinated. 
 
 The results of prevalence estimation via the methods developed in Section \ref{sec:prev_est} (not shown) are essentially identical: the norms of the differences, taken across all time periods, of the means and standard deviations produced by the two methods are less than $1.5 \times 10^{-14}$ for both $I$ and $V$. This is as expected, as we have shown in Subsection \ref{sec:equiv_meas_dens} that the measurement densities under these two frameworks are equivalent. 

\section{Discussion}
\label{sec:disc}

We have shown the equivalence of the probabilistic modeling approaches in \cite{bedekar2022prevalence} and \cite{luke2023optimal} and the time-inhomogeneous Markov chain approach for the simple case of multiclass time-dependent effects that preclude reinfection, revaccination, or any cross-effects. Useful immediately after  the introduction of vaccines, this case is restrictive for general use. The equivalence itself is a significant contribution as we anticipate that our Markov chain approach will be effective in modeling the most general case. As a very promising result, prevalence estimation via transition probability matrices is identical in expectation to that of the multiclass extension of the existing time-dependent framework (Section \ref{sec:methods}); the former can reliably be used as we address the most general version of the problem.

\subsection{More on prevalence estimation}
\label{sec:disc_prev_est}

We begin with a few comments on prevalence estimation. In Section \ref{sec:prob_models}, we noted how we discretize time in time steps of $dt$ days to follow batched reporting trends. Daily measurements are rarely available and testing delays occur; using a larger value of $dt$ provides less detailed information, but results in better prevalence estimates when data are sparse \citep{bedekar2022prevalence}. We also note that for prevalence estimation via transition probability matrices, the system given by Eq. \eqref{eq:tranProbEst0} is invertible provided the choice of  time step $dt$ and subdomain partition is such that $R_{D_j}(1), W_{D_j}(1), \text{ and } N_{D_j}$ are well separated from each other. This ensures that the antibody response of infected or vaccinated individuals is separable from that of na{\"i}ve population. Further, a careful selection of subdomains $D_j$ as guided by \cite{patrone2024minimizing}, \cite{luke2023optimal}, and \cite{bedekar2022prevalence} can help the estimation by minimizing numerical errors.

\begin{figure}[ht]
\centering
\subfloat[][Overlapping]{\includegraphics[scale=0.4]{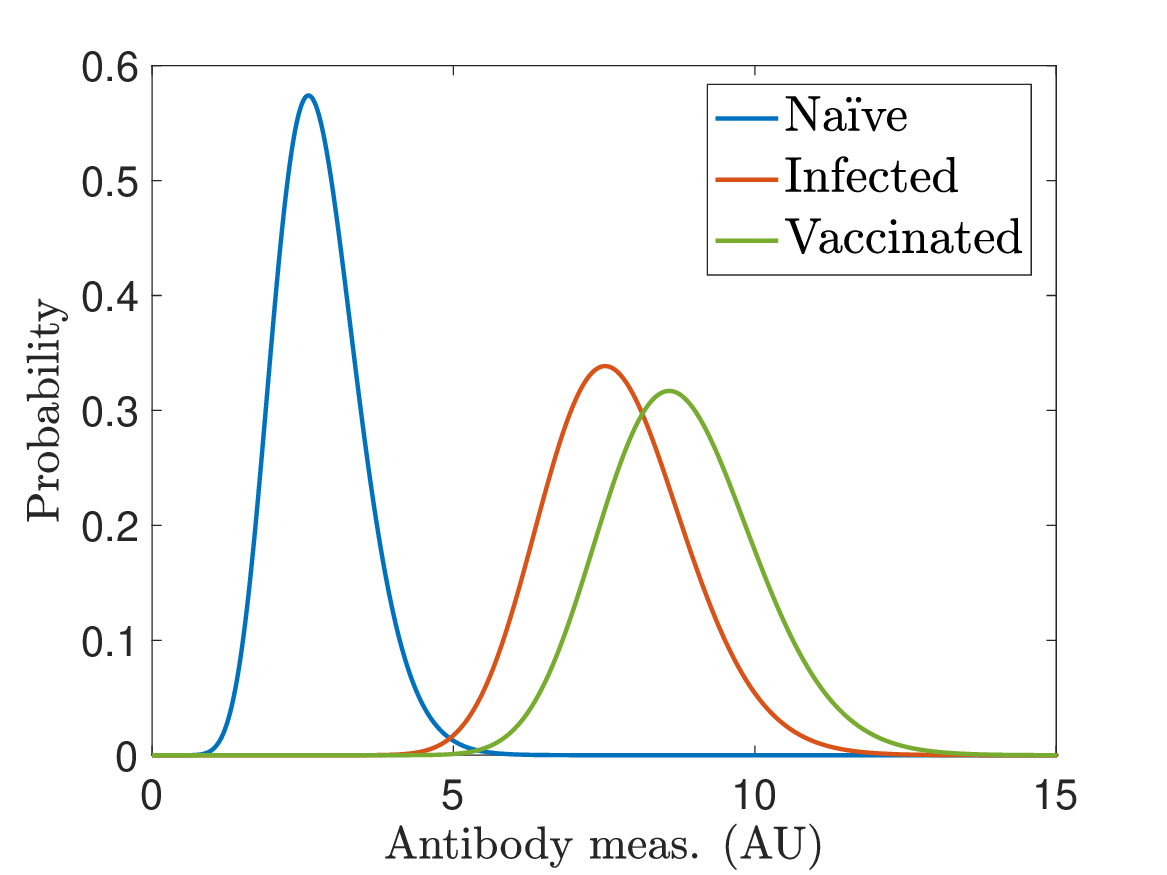}}
\subfloat[][Separated]{\includegraphics[scale=0.4]{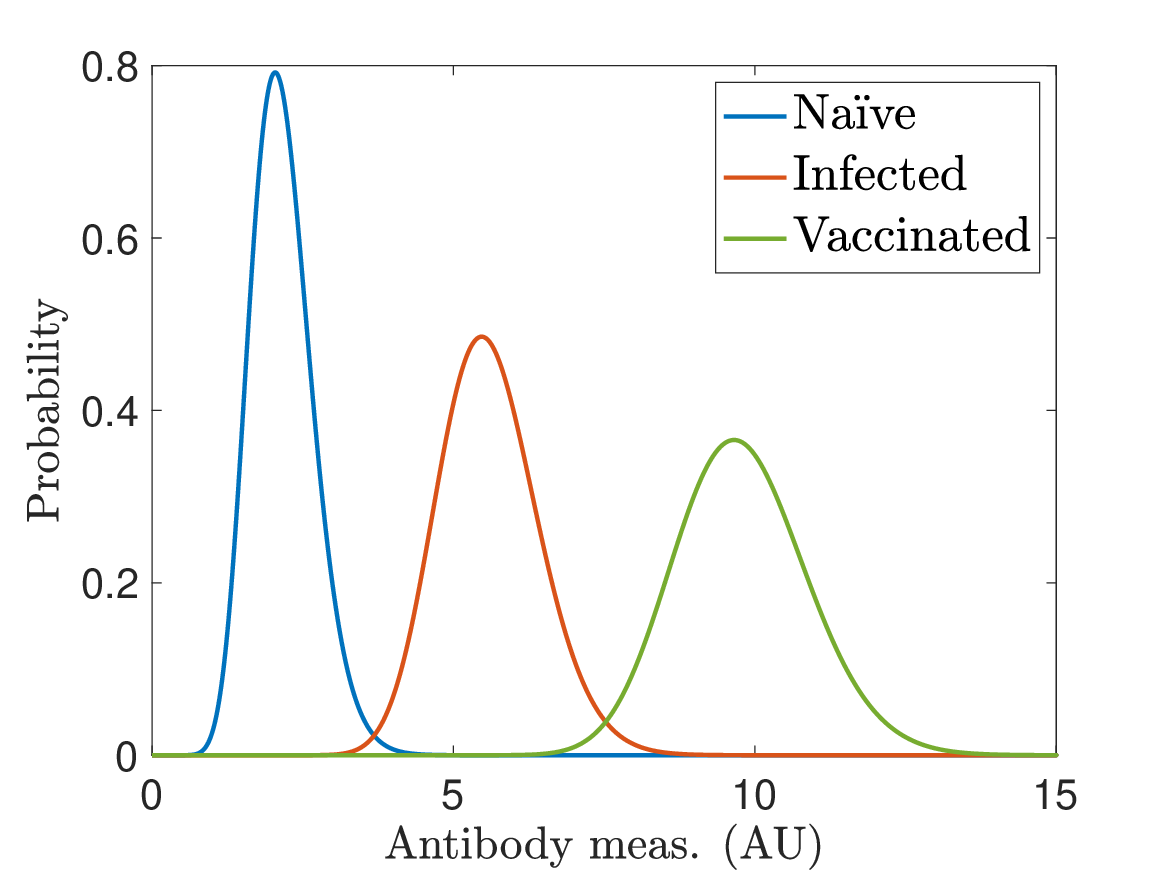}}
\caption{Distributions of antibody responses at time step 1 (21 days) for partially overlapping synthetic data motivated by \cite{abela2021multifactorial} and \cite{congrave2022twelve} and synthetic data artificially generated to be more separated at $t = 21$.}
\label{fig:prev_est_dist}
\end{figure}

\begin{figure}[ht]
\centering
\subfloat[][Wave for previously infected]{\includegraphics[scale=0.3]{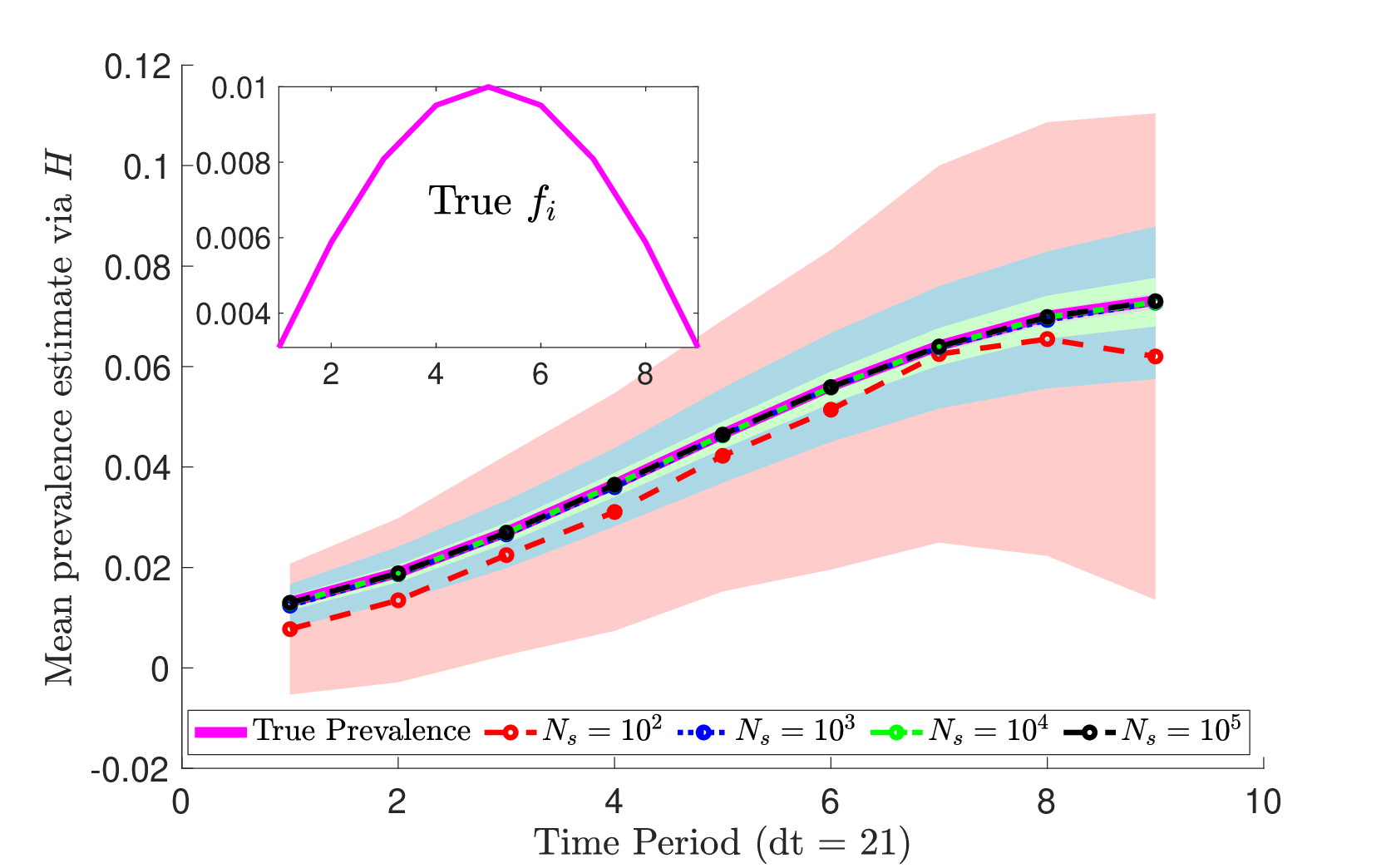}}
\subfloat[][Constant newly vaccinated]{\includegraphics[scale=0.3]{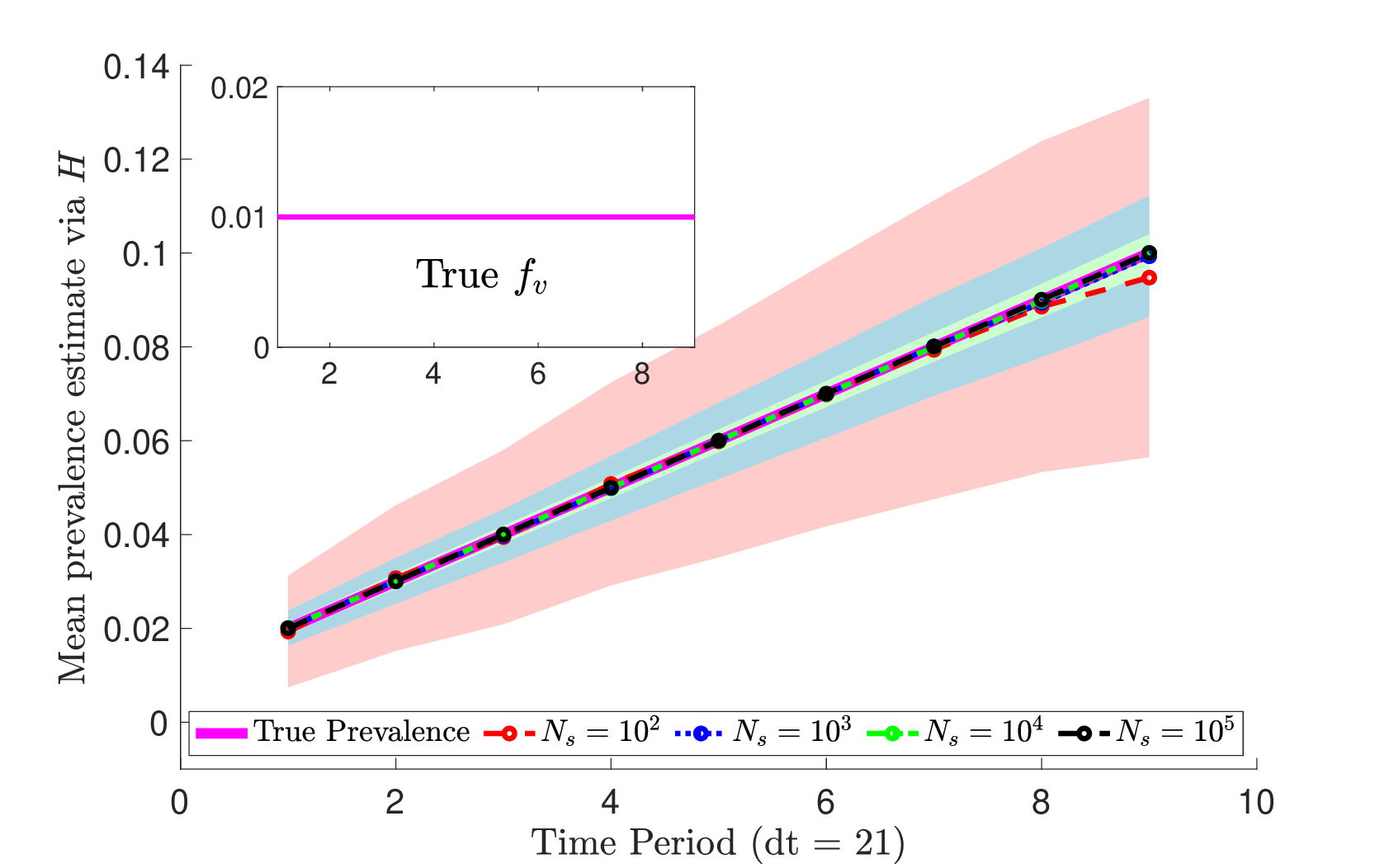}}
    \caption{Prevalence estimation via transition probability estimates using synthetic data for antibody responses. The mean over 1000 synthetic data sets is shown for various numbers of samples $N_s$ with standard deviation confidence intervals (shown in a lighter shade of the corresponding color of $N_s$) over time.}
    \label{fig:prev_est_H_sep}
\end{figure}

 We now revisit prevalence estimation via transition probability matrices after observing the large errors and standard deviations in our example shown in Figure \ref{fig:prev_est_H_not_sep}. A known issue in prevalence estimation and classification is overlap of class data \citep{luke2023optimal}. By plotting the probability models at one time step (21 days) in Figure \ref{fig:prev_est_dist}a, we note that the previously infected and vaccinated distributions exhibit significant overlap. As an exercise, we artificially create populations that exhibit more separation at the first time step using similar gamma distributions\footnote{ The model parameters are $\alpha_n = 17.6, \beta_n = 0.123$, $\theta_{1, i} = 2.23, \theta_{2, i} = 5.3 \times 10^{-4}, \theta_{1, v} = 4.23$, and $\theta_{2, v} = 5.3 \times 10^{-4}$.}, as shown at time step 1 in Figure \ref{fig:prev_est_dist}b.  We conduct prevalence estimation via transition probability matrices for test data sampled from these distributions  and display the results in Figure \ref{fig:prev_est_H_sep}. As before,  prevalence estimation  using the methods developed in Section \ref{sec:prev_est} yields essentially identical results due to the equivalence in measurement densities, as proven in Subsection \ref{sec:equiv_meas_dens}.

 Note that a smaller number of samples, $N_s = 10^2$, is shown in Figure \ref{fig:prev_est_H_sep} as compared to Figure \ref{fig:prev_est_H_not_sep}, and there are no negative prevalence estimates observed by using $N_s = 10^4$.  Further, by using  $N_s = 10^4$ samples, we are able to achieve average prevalence estimate errors across all time periods of $(5.6 \pm 4.6)$ \% for  infected and $(3.6 \pm 2.8)$ \% for vaccinated, which is significantly better than the errors shown in Figure \ref{fig:prev_est_H_not_sep} for $N_s = 10^5$. Using $N_s = 10^5$ for these better separated populations improves our errors to $(1.9 \pm 1.5)$ \% for  infected and $(1.2 \pm 0.9)$ \% for vaccinated.

\begin{figure}[ht]
\centering
\includegraphics[scale=0.55]{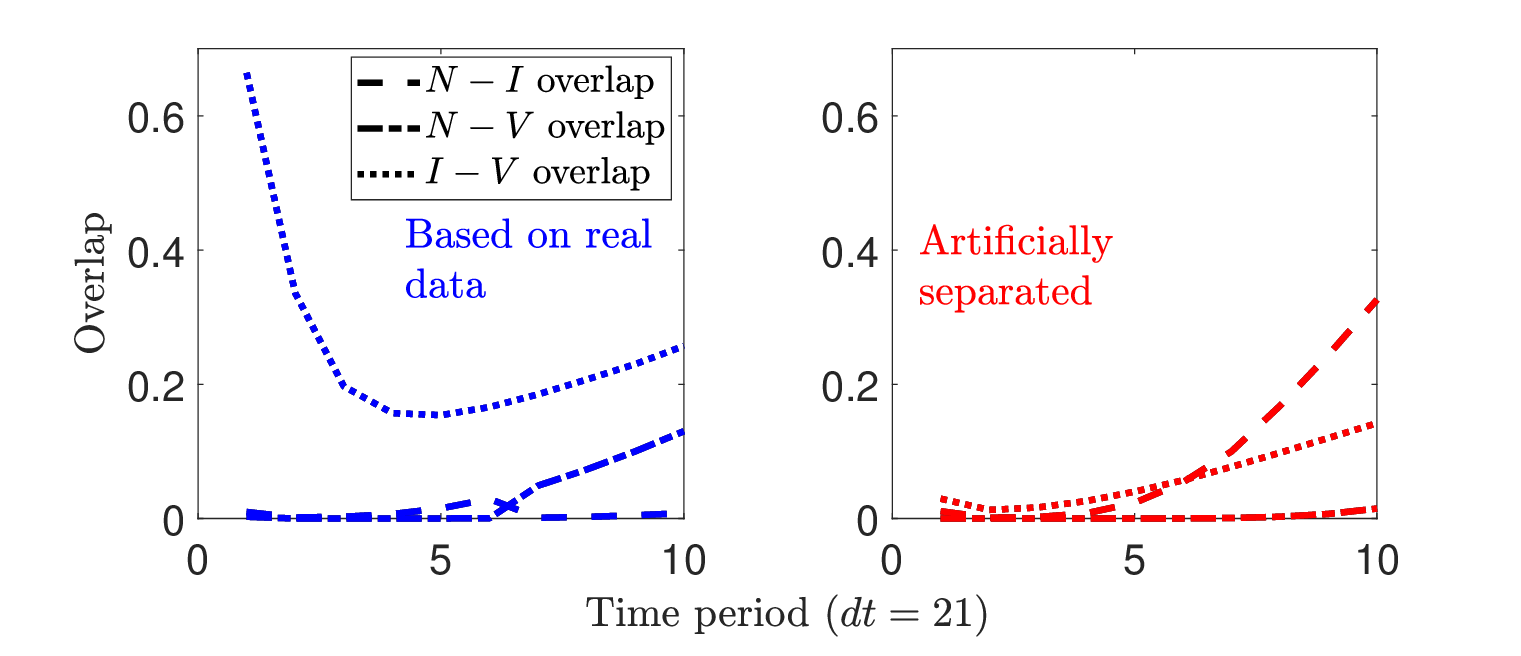}
\caption{Measure of overlap of the antibody responses over time for the models that generated Figure \ref{fig:prev_est_dist}.}
\label{fig:overlap}
\end{figure}

How many sample points are needed from the population at each time step might vary based on how separated the populations are, which can change over time. The importance of sampling may depend on the disease in question, which determines how separated the populations are naturally. We investigate the degree of separation over time of the distributions from the above example using the overlap metric defined by \cite{weitzman1970measures} as the shared area under two probability densities:

 \begin{equation}
     \text{OVL}(X,Y) = \int_{-\infty}^{\infty} \min \{ P_1(x), P_2(x) \} \ dx.
 \end{equation}
 We compare the pairwise overlap over time in Figure \ref{fig:overlap}. The overlapping  infected and vaccinated distributions (Figure \ref{fig:prev_est_dist}a) agree on two-thirds of their underlying area at time step 1, whereas the artificially separated distributions have 3 \% agreement at the same time. Interestingly, at time step 10 there is greater overlap of the infected and vaccinated populations for the artificially separated data; this suggests that separability is crucial in the early time steps due to the accumulation of errors. \cite{schmid2006nonparametric} provide nonparametric estimators for the coefficient of overlap that could provide insight into the expected difficulty of prevalence estimation given sample populations with high levels of similarity. In future work, ideas from these authors could be extended to identify the number of samples needed to perform prevalence estimation with low variance. 

For this synthetic example, we have demonstrated that better separation of populations in the first few time steps can result in more accurate prevalence estimation. We now discuss how to separate real data.
First and foremost, if additional measurements per person are available, such as values for other protein markers, one can visualize the data in higher dimensions, thereby inducing separation \citep[see e.g.,][]{luke2023improving}. If additional measurements are not available, there are ways to embed the data into a higher dimension, such as basis expansion from the field of kernel methods \citep{hastie2009elements}. Metrics like the silhouette coefficient for data and the KL-divergence for distributions indicate a similarity score between populations, which we expect can determine if the populations are separated ``enough'' to conduct prevalence estimation.
One can also borrow the idea of a holdout domain from \cite{patrone2022holdout} to exclude highly overlapped data from the prevalence estimation. 
Instead of a multiclass prevalence estimation, a two level binary procedure can be conducted: first to estimate the relative sizes of the na{\"i}ve and not-na{\"i}ve populations, and then another  to separate the previously infected and vaccinated classes. For the secondary step, holdout analysis or other separation methods can be applied as needed.
Finally, one can compute an altered estimate $\hat{\bm{Q}}$ by summing indicator functions weighted inversely proportional to the overlap of the populations. This may lose the desirable property of unbiasedness, but we expect this to be overlooked in favor of reasonable prevalence estimates.

\subsection{Relationship to susceptible--infected--recovered models}

Our approach is superior to an SIR framework, which can model disease transmission but cannot track the distribution of antibody response of a population across time. In contrast, we provide such probabilistic information as well as a prevalence estimation scheme that is independent of classification. However, there are connections between the two approaches. Probabilistic descriptions of SIR models exist, including Markov chain formulations \citep[e.g.,][]{liao2005probabilistic,cortes2020comprehensive,el2021mathematical}. Within-host SIR-like deterministic versions explicitly account for antibody densities and viral load as variables \citep{hancioglu2007dynamical}. Our na{\"i}ve state may be a proxy for the susceptible compartment, with a population-level equivalence of these two categories. We noted in Section \ref{sec:prob_models} that unlike an SIR model, we have no recovered class; this leads to an interesting question. It is unclear how to mathematically describe the process by which someone becomes na{\"i}ve after  previous infection or vaccination. Our current framework assumes equivalence in the limit of time, but for a seasonal disease, there should be a nonzero probability of returning to $N$ in finite time. In the language of Markov chains, we speculate that the na{\"i}ve state is a recurrent state in finite time for the graph model that allows transition back to na{\"i}ve state.

 \subsection{Limitations and extensions}

 The choice of the form of the na{\"i}ve model and shape functions for the infected and vaccinated populations are subjective choices \citep{smith2013uncertainty}, but the influence of this issue is lessened as more sample points are used \citep{schwartz1967estimation}. A family of models may be proposed and the one selected with minimal error on a measure of interest \citep{patrone2024minimizing}, such as prevalence estimates. As in our previous works, we have noted that the overlap of populations increases prevalence estimation error; we have provided potential solutions in Section \ref{sec:disc_prev_est}. Additionally, \cite{bedekar2022prevalence} first noted that prevalence estimation errors accumulate over time; our current scheme does not address controlling such errors.

 Perhaps the largest drawbacks to our current approach are the simplifying assumptions we make to construct the groundwork for a multiclass time-dependent framework. For long-term analysis of disease emergence effects, multiple events must be allowed. 
In future work, we will relax the assumption that reinfection, revaccination, and infection after vaccination or vice versa do not occur. 
The many facets of the most general problem make tracking the immune response difficult, but we must consider such cases, because reinfections and revaccinations are the norm as a newly emergent disease becomes endemic.
There is little biological understanding of interactions between such events and precious few models, as the antibody kinetics are still being studied. 
Modeling how these effects ``stack'' on each other in terms of immune response is a complex question. However, there are some guiding principles from which to start.  Extreme sequences of events, such as an individual becoming newly infected every day, should be assigned very low probabilities due to the underlying biology. Further, despite the multitude of potential sequences of events leading to an antibody response and current state on a particular time step, the likelihood of infection or vaccination on the next time step depends solely on the current state information. Additionally, the multi-step transition probability matrix in the current graph framework is considerably more structured than can be expected from the more general model. As a result, the estimation methods will need to be revised. These and other considerations inform our ongoing work to form a general model that can tackle real-life scenarios. 

There are significant data-related challenges that should be addressed in future work. One is simply a lack of temporal data, which affects the modeling process and prevalence estimation. Even when data are available, it can poorly represent the progression of an emerging disease through a population as truly random testing is not realistic.
 In the case of SARS-CoV-2, it is well known that people tested less frequently as the pandemic progressed, false negative tests occurred, and people often did not recognize mild or asymptomatic cases as infections. Uncertainty quantification could address the deterioration of accuracy and precision of collected antibody data since the start of the pandemic. Prevalence estimation is a data-dependent analysis, which begins by estimating the incidence of new infections and vaccinations. Vaccination incidence rates may be well-documented for a population if de-identified medical data are available, but new infection case rates are prone to missing responses and errors due to the inexactness of days post symptom onset (DPSO) as an infection marker. Moreover, DPSO may often understimate the true time since the beginning of infection. We note that immunocompromised individuals affect prevalence estimation and the modeling exercise. Future  work could address the propagation of error by separating the population by immune system status, and correct errors due to reporting bias or gaps. We could also consider inflammatory markers as candidate variables, which have been studied in the context of severity of diseases such as the coronavirus disease of 2019 (COVID-19) \citep{zeng2020association}, but may not be fully understood in connection with sequences of immune response events.

 \subsection{Implications for immunologists}

We have created a cohesive framework for the multiclass time-dependent problem of the emergence of a disease, so that unbiased predictions of the relative fractions of na{\"i}ve, infected, and vaccinated individuals can be generated over time. Although we use SARS-CoV-2 as a motivating example, this approach is fully generalizable to other diseases for which immunity is lost on the time frame of months or a few years. In particular, the models follow biological assumptions that can be adapted or narrowed to focus on populations of interest, such as children, the elderly, or the immunocompromised.  Our methods are limited by data availability; we recommend implementing longitudinal studies that continue to record infections with high granularity even as vaccines are deployed. As assay standardization is not fully achieved, such studies should use the same data collection methods, instruments, and protocol to facilitate the comparison of measurements across large periods of time.

\section*{CRediT authorship contribution statement}

\textbf{Prajakta Bedekar}: Conceptualization, Data curation, Formal analysis, Methodology, Software, Validation, Visualization, Writing--original draft, Writing--review \& editing. \textbf{Rayanne A. Luke}: Conceptualization, Data curation, Formal analysis, Methodology, Software, Validation, Visualization, Writing--original draft, Writing--review \& editing. \textbf{Anthony J. Kearsley}: Conceptualization, Methodology, Supervision, Writing--review \& editing.

 \section*{Declaration of competing interest} The authors declare that they have no known competing financial interests or personal relationships that could have appeared to influence the work reported in this paper.

 \section*{Data availability} Analysis scripts and data developed as a part of this work are available upon reasonable request. Motivating data are provided in \cite{abela2021multifactorial} and \cite{congrave2022twelve}.

 \section*{Acknowledgements} This work is a contribution of the National Institute of Standards and Technology, USA and is not subject to copyright in the United States. P.B. was funded through the NIST PREP, USA grant 70NANB18H162. R.L. is grateful for generous support during the early days of the research from a National Research Council postdoctoral fellowship at NIST and a NIST PREP fellowship (same grant number as P.B.).  The aforementioned funders had no role in study design, data analysis, decision to publish, or preparation of the manuscript. The authors are grateful to Pia Pannaraj and Yesun Lee for helpful discussion regarding post-infection or vaccination antibody kinetics. The authors also wish to thank Melinda Kleczynski and Amanda Pertzborn for useful feedback during preparation of this manuscript.

\bibliographystyle{elsarticle-harv}

\bibliography{bibliography.bib}

\setcounter{equation}{0}
\setcounter{figure}{0}
\setcounter{table}{0}
\appendix
\renewcommand{\theequation}{A\arabic{equation}}
\renewcommand{\thefigure}{A\arabic{figure}}
\renewcommand{\thetable}{A\arabic{table}}

 \section{Unbiasedness of the prevalence estimators}
 \label{sec:unbiased}

    A straightforward extension of the ideas in \cite{bedekar2022prevalence} shows that the prevalence estimates presented in Section \ref{sec:prev_est} are unbiased. In what follows we continue to use element-wise vector addition to simplify our notation. First, it is clear that $\hat{\bm{Q}}(T)$ is unbiased because each component is a Monte Carlo estimator and therefore 
    
    \begin{equation}
        \hat{Q}_j(T) \sim \frac{1}{S} \text{Binomial}(S, Q_j(T)),\label{eq:unbiasedQ}\end{equation} which implies 
        
    \begin{equation}
        E\left[\hat{Q}_j(T) \right] = \frac{1}{S} E \left[ \text{Binomial}(S, Q_j(T))\right] = Q_j(T).
    \end{equation}
   Next, we find that
   
    \begin{equation}
       E \left[ \hat{\bm{f}}(0) \right] = E \left\{  [\bm{M}(1) - \bm{N}^*]^{-1} [\bm{\hat{Q}}(1) - \bm{N}]\right\} =  [\bm{M}(1) - \bm{N}^*]^{-1} [E \{\bm{\hat{Q}}(1) \} - \bm{N}],
    \end{equation}
    and using the previous result we conclude that $E \left[ \hat{\bm{f}}(0) \right] = \bm{f}(0)$. Then, via induction, we find that
    
\begin{equation}
\begin{split}
E \left[ \hat{\bm{f}}(T-1) \right] & = E \left\{ [\bm{M}(1) - \bm{N}^*]^{-1} \left[  \bm{\hat{Q}}(T) - \bm{N} - \sum_{t = 0}^{T-2} [\bm{M}(T-t) - \bm{N}^*] \bm{\hat{f}}(t) \right] \right\} \\
& = [\bm{M}(1) - \bm{N}^*]^{-1} \left[ E \left[\bm{\hat{Q}}(T) \right] - \bm{N} - \sum_{t = 0}^{T-2}  [\bm{M}(T-t) - \bm{N}^*] E \left[ \bm{\hat{f}}(t) \right] \right] \\
& = [\bm{M}(1) - \bm{N}^*]^{-1} \left[  \bm{Q}(T) - \bm{N} - \sum_{t = 0}^{T-2}  [\bm{M}(T-t) - \bm{N}^*] \bm{f}(t) \right] \\
& = \bm{f}(T-1).
\end{split}
\end{equation}
Finally, we find that

\begin{equation}
E \left[ \hat{\bm{q}}(T) \right] = E \left[ \sum_{t = 0}^T \hat{\bm{f}}(t) \right] = \sum_{t = 0}^T E \left[ \bm{\hat{f}}(t) \right] = \sum_{t = 0}^T \bm{f}(t) = \bm{q}(T).
\end{equation}

\setcounter{equation}{0}
\setcounter{figure}{0}
\setcounter{table}{0}
\renewcommand{\theequation}{B\arabic{equation}}
\renewcommand{\thefigure}{B\arabic{figure}}
\renewcommand{\thetable}{B\arabic{table}}

\section{Unbiasedness of transition probability estimation}
      \label{sec:unbiased_trans}

The proof of the unbiasedness of the transition probability matrix entries can be obtained easily by using similar techniques as in \ref{sec:unbiased}. We use linearity of expectation of the product of a deterministic matrix and  a random vector on Eq. \eqref{eq:tranProbEst0} to obtain 

\begin{equation}
     \hspace{-10mm}  E\left(\begin{bmatrix}
        \widehat{q}_N(0) \\ \widehat{H}_{0,(3,1)}\\ \widehat{H}_{0,(5,1)}
    \end{bmatrix}\right) = E\left(\begin{bmatrix}
N_{D_1} & R_{D_1}(1) & W_{D_1}(1)\\
N_{D_2} & R_{D_2}(1) & W_{D_2}(1)\\
        1 & 1 & 1
    \end{bmatrix} ^{-1}\begin{bmatrix}
        \widehat{Q}_{D_1}(1) \\ \widehat{Q}_{D_2}(1)\\ 1
    \end{bmatrix} \right) = \begin{bmatrix}
N_{D_1} & R_{D_1}(1) & W_{D_1}(1)\\
N_{D_2} & R_{D_2}(1) & W_{D_2}(1)\\
        1 & 1 & 1
    \end{bmatrix} ^{-1}E\left(\begin{bmatrix}
        \widehat{Q}_{D_1}(1) \\ \widehat{Q}_{D_2}(1)\\ 1
    \end{bmatrix} \right).
\end{equation}
Using the unbiasedness of the Monte Carlo estimates as shown in Eqs. \eqref{eq:unbiasedQ} and \eqref{eq:tranProbEst0} again, we get

\begin{equation}
\label{eq:unbiasedTranProb0}       E\left(\begin{bmatrix}
        \widehat{q}_N(0) \\ \widehat{H}_{0,(3,1)}\\ \widehat{H}_{0,(5,1)}
    \end{bmatrix}\right) = \begin{bmatrix}
N_{D_1} & R_{D_1}(1) & W_{D_1}(1)\\
N_{D_2} & R_{D_2}(1) & W_{D_2}(1)\\
        1 & 1 & 1
    \end{bmatrix} ^{-1} \begin{bmatrix}
        Q_{D_1}(1) \\ Q_{D_2}(1)\\ 1
    \end{bmatrix} = \begin{bmatrix}
        q_N(0) \\ H_{0,(3,1)}\\ H_{0,(5,1)}
    \end{bmatrix}.
\end{equation}
These calculations are easily generalized by using the principle of strong induction and Eqs. \eqref{eq:tranProbEstT}, \eqref{eq:unbiasedQ}, and \eqref{eq:unbiasedTranProb0},
to obtain

\begin{equation}
\hspace{-20mm}\begin{split}
\label{eq:unbiasedTranProbT}
        E\left(\begin{bmatrix}
        \widehat{q}_N(T-1) \\ \widehat{H}_{(T-1)(3,1)}\\ \widehat{H}_{(T-1)(5,1)}
    \end{bmatrix}\right)
    & = \begin{bmatrix}
        N_{D_1} & R_{D_1}(1) & W_{D_1}(1)\\
        N_{D_2} & R_{D_2}(1) & W_{D_2}(1)\\
        1 & 1 & 1
    \end{bmatrix} ^{-1} E\left(\begin{bmatrix}
    \widehat{Q}_{D_1}(T) - \sum\limits_{t=0}^{T-2} \left( R_{D_1}(T-t) \widehat{H}_{t,(3,1)} + W_{D_1}(T-t) \widehat{H}_{t,(5,1)} \right)\\ \widehat{Q}_{D_2}(T) - \sum\limits_{t=0}^{k-2} \left( R_{D_2}(T-t) \widehat{H}_{t,(3,1)} + W_{D_2}(T-t) \widehat{H}_{t,(5,1)} \right)\\ \widehat{q}_N(T-2)
    \end{bmatrix}\right)\\
    & = \begin{bmatrix}
        N_{D_1} & R_{D_1}(1) & W_{D_1}(1)\\
        N_{D_2} & R_{D_2}(1) & W_{D_2}(1)\\
        1 & 1 & 1
    \end{bmatrix}^{-1} \begin{bmatrix}
        Q_{D_1}(T) - \sum\limits_{t=0}^{T-2} \left( R_{D_1}(T-t) H_{t,(3,1)} + W_{D_1}(T-t) H_{t,(5,1)} \right)\\ Q_{D_2}(T) - \sum\limits_{t=0}^{T-2} \left( R_{D_2}(T-t) H_{t,(3,1)} + W_{D_2}(T-t) H_{t,(5,1)} \right)\\ q_N(T-2)
    \end{bmatrix}\\
     & = \begin{bmatrix}
        q_N(T-1) \\ H_{(T-1)(3,1)}\\ H_{(T-1)(5,1)}
    \end{bmatrix}.
    \end{split}
\end{equation}
The estimates for the other entries of the transition probability matrix are also unbiased. Using Eqs. \eqref{eq:tranProbH2} and \eqref{eq:tranProbH4} we see that

\begin{equation}
E\left(\widehat{H}_{(T-1)(2,1)}\right) =  \sum\limits_{\tau=0}^{T-2} E\left(\widehat{H}_{\tau(3,1)}\right) = \sum\limits_{\tau=0}^{T-2} H_{\tau(3,1)} = H_{(T-1)(2,1)}.
\end{equation}
and 

\begin{equation}
E\left(\widehat{H}_{(T-1)(4,1)}\right) = \sum\limits_{\tau=0}^{T-2} E\left(\widehat{H}_{\tau(5,1)}\right)
= \sum\limits_{\tau=0}^{T-2} H_{\tau(5,1)}
 = H_{(T-1)(4,1)}.
 \end{equation}

\section{Optimal classification}
      \label{sec:classif}

Optimal classification can be performed to yield class domains that vary over time. Since the antibody kinetics of states $N$ and $\pos$ share the same distribution, newly infected individuals cannot be distinguished from those na{\"i}ve to the disease. This reflects the known delay in antibody response after infection \citep{borremans2020quantifying}. Thus, antibody measurements are classified as belonging to one of  $N$, $\posp$, or $\vacp$.

We combine the ideas in \cite{bedekar2022prevalence} and \cite{luke2023optimal} to construct the optimal classification domains that minimize a loss function. We seek to define a sequence of partitions $\{D_N(T), D_{\posp}(T), D_{\vacp}(T)\}$, not necessarily the same as any partition used for prevalence estimation, so that at any time $T$, a measurement is assigned to one and only one class, i.e., $N$ if $\bm{r} \in D_N(T)$, and similarly for $D_{\posp}(T)$ and $D_{\vacp}(T)$. To ensure any sample can be classified and to enforce single-label classification up to sets of measure zero, we require, for any $T$, that
    \begin{eqnarray}
        \mu_N \left(\cup_D \right) = \mu_{\posp} \left( \cup_D \right) = \mu_{\vacp} \left( \cup_D \right) = 1,
        \label{eq:rule1}\\
        \mu_{J} [D_{L}(T) \cap D_K(T) ] = 0 \text{ for } L \neq K, \text{ for } J \in \{N, \posp, \vacp \}.
        \label{eq:rule2}
    \end{eqnarray}
    Here, $\cup_D =  D_N(T) \cup D_{\posp}(T) \cup D_{\vacp}(T)$, the subscripts $L$ and $K$ can represent any of $N, \posp$, or $\vacp$, and $\mu_N(X) = \int_X N(\bm{r}) d \bm{r}$, $\mu_{\posp}(X) = \int_X \posp(\bm{r}, T) d \bm{r}$, and $\mu_{\vacp}(X) = \int_X \vacp(\bm{r}, T) d \bm{r}$.

    Next, define the prevalence-weighted rate of misclassification at time $T$:
    \begin{dmath}
     \mathscr{L}\left(D_N(T), D_{\posp}(T), D_{\vacp}(T) \right)  =   q_N(T) \int_{\Omega \setminus D_N(T)} N(\bm{r}) d \bm{r} + q_{\posp}(T) \int_{\Omega \setminus D_{\posp}(T)} \posp(\bm{r}, T) d \bm{r} + q_{\vacp}(T) \int_{\Omega \setminus D_{\vacp}(T)} \vacp(\bm{r}, T) d \bm{r}.
    \end{dmath}
    One expects that a sample $\bm{r}$ should be assigned to the class to which it has the highest probability of belonging on time step $T$; that is, $\max \{ q_N(T) N(\bm{r}) , q_{\posp}(T) \posp(\bm{r}, T), q_{\vacp}(T) \vacp(\bm{r}, T) \}$.  It turns out that our intuition is correct. 
    The authors in \cite{luke2023optimal} showed that for a single time $T$, the optimal classification domains are given by
    \begin{subequations}
    \begin{eqnarray}
        D_N^{\star}(T) = \{ \bm{r} : q_N(T) N(\bm{r}) > q_{\posp}(T) \posp(\bm{r}, T) \} \cap \{\bm{r}:  q_N(T) N(\bm{r}) > q_{\vacp}(T) \vacp(\bm{r},T)\}, \\
        D_{\posp}^{\star}(T) = \{ \bm{r} : q_{\posp}(T) \posp(\bm{r}, T) > q_N(T) N(\bm{r}) \} \cap \{\bm{r}:  q_{\posp}(T) \posp(\bm{r}, T) > q_{\vacp}(T) \vacp(\bm{r},T)\}, \\
        D_{\vacp}^{\star}(T) = \{ \bm{r} : q_{\vacp}(T) \vacp(\bm{r}, T) > q_N(T) N(\bm{r}) \} \cap \{\bm{r}:  q_{\vacp}(T) \vacp(\bm{r}, T) > q_{\posp}(T) \posp(\bm{r},T)\}.
        \end{eqnarray}
        \label{eq:opt_pw}
    \end{subequations}
   \hspace{-2.5mm} This assumes a technical detail that all ``boundary'' cases, i.e., measurements $\bm{r}$ such that there is an equal highest probability of belonging to two or more classes, have measure zero. Since this is often true for any practical implementation, we will assume this going forward. Also note that the use of a superscript $\star$ denotes an optimal quantity, in this case, an optimal classification domain.
    
    A logical loss function is then the sum of these rates over all time steps of the emergence of the disease:
    
    \begin{equation}
        \mathscr{L}_{\tau} (\bm{D}_{\bm{N}}, \bm{D}_{\bm{\posp}}, \bm{D}_{\bm{\vacp}}) = \sum_{T = 0}^{\tau} \mathscr{L}(D_N(T), D_{\posp}(T), D_{\vacp}(T)),
        \label{eq:loss}
    \end{equation}
    where each of $D_N(T), D_{\posp}(T), D_{\vacp}(T)$ obey the rules given by Eqs. \eqref{eq:rule1}-\eqref{eq:rule2} and $\bm{D}_{\bm{N}}$ is a vector whose $i$th entry is $D_N(i)$. The vectors $\bm{D}_{\bm{\posp}}$ and $\bm{D}_{\bm{\vacp}}$ are defined analogously. A straightforward application of the ideas in \cite{bedekar2022prevalence} shows us that the loss function \eqref{eq:loss} is minimized by taking the pointwise-optimal domains at each time step $T$ given by Eq. \eqref{eq:opt_pw}. Thus, the optimal classification domains are given by the vectors 
    
    \begin{equation}
    \bm{D}_{\bm{N}}^{\star} = \begin{bmatrix}
            D_N^{\star}(0) \\
            D_N^{\star}(1) \\
             \vdots \\
             D_N^{\star}(\tau)
        \end{bmatrix}, \quad \bm{D}_{\bm{\posp}}^{\star} = \begin{bmatrix}
            D_{\posp}^{\star}(0) \\
            D_{\posp}^{\star}(1) \\
             \vdots \\
             D_{\posp}^{\star}(\tau)
        \end{bmatrix}, \quad \bm{D}_{\bm{\vacp}}^{\star} = \begin{bmatrix}
            D_{\vacp}^{\star}(0) \\
            D_{\vacp}^{\star}(1) \\
             \vdots \\
             D_{\vacp}^{\star}(\tau)
        \end{bmatrix}.
    \end{equation} 

\end{document}